\documentclass[sigconf, nonacm]{acmart}

\newcommand\vldbavailabilityurl{https://github.com/starkersawz666/KATS}
 
\newcommand{\systemname}{KATS}
\newcommand{\benchmarkname}{CS-TDS}

\usepackage{booktabs}
\usepackage{enumitem}
\usepackage{tabularx}
\usepackage{multirow}
\usepackage{minted}
\usepackage{fancyvrb}
\usepackage{listings}
\usepackage{xcolor}
\usepackage{tcolorbox}
\usepackage{cleveref}
\tcbuselibrary{skins}
\usepackage{cuted}
\usepackage{balance}

\newcommand{\revised}[1]{{\color{black}#1}}

\lstdefinestyle{jsonstyle}{
    language=json,
    basicstyle=\small\ttfamily,
    stringstyle=\color{purple},
    keywordstyle=\color{blue},
    numberstyle=\tiny\color{gray},
    commentstyle=\color{gray},
    showstringspaces=false,
    breaklines=true,
    frame=tb,
    backgroundcolor=\color{black!5},
    literate=
      *{0}{{{\color{gray}0}}}{1}
       {1}{{{\color{gray}1}}}{1}
       {2}{{{\color{gray}2}}}{1}
       {3}{{{\color{gray}3}}}{1}
       {4}{{{\color{gray}4}}}{1}
       {5}{{{\color{gray}5}}}{1}
       {6}{{{\color{gray}6}}}{1}
       {7}{{{\color{gray}7}}}{1}
       {8}{{{\color{gray}8}}}{1}
       {9}{{{\color{gray}9}}}{1}
}

\begin{document}
\title{Revisiting Task-Oriented Dataset Search in the Era of Large Language Models: Challenges, Benchmark, and Solution}

\author{Zixin Wei}
\affiliation{%
  \institution{The Chinese University of Hong Kong, Shenzhen}
  \city{Shenzhen}
  \country{China}
}
\email{zixinwei1@link.cuhk.edu.cn}

\author{Yucan Guo}
\authornote{Also affiliated with School of Computer Science and Technology, University of Chinese Academy of Sciences.}
\affiliation{%
  \institution{Institute of Computing Technology, \\Chinese Academy of Sciences}
  \city{Beijing}
  \country{China}}
\email{guoyucan23z@ict.ac.cn}

\author{Jinyang Li}
\affiliation{%
  \institution{The University of Hong Kong}
  \city{Hong Kong}
  \country{China}}
\email{jl0725@connect.hku.hk}

\author{Xiaolin Han}
\affiliation{%
  \institution{The Northwestern Polytechnical University}
  \city{Xi'an}
  \country{China}}
\email{xiaolinh@nwpu.edu.cn}

\author{Xiaolong Jin}
\authornotemark[1]
\affiliation{%
  \institution{Institute of Computing Technology, \\Chinese Academy of Sciences}
  \city{Beijing}
  \country{China}}
\email{jinxiaolong@ict.ac.cn}

\author{Chenhao Ma}
\authornote{Corresponding author.}
\affiliation{%
  \institution{The Chinese University of Hong Kong, Shenzhen}
  \city{Shenzhen}
  \country{China}
}
\email{machenhao@cuhk.edu.cn}

\begin{abstract}
The search for suitable datasets is the critical ``first step'' in data-driven research, but it remains a great challenge. Researchers often need to search for datasets based on high-level task descriptions. However, existing search systems struggle with this task due to ambiguous user intent, task-to-dataset mapping and benchmark gaps, and entity ambiguity. To address these challenges, we introduce \systemname{}, a novel end-to-end system for task-oriented dataset search from unstructured scientific literature. \systemname{} consists of two key components, i.e., offline knowledge base construction and online query processing. 
The sophisticated offline pipeline automatically constructs a high-quality, \revised{dynamically updatable} task-dataset knowledge graph by employing a collaborative multi-agent framework for information extraction, thereby filling the task-to-dataset mapping gap. To further address the challenge of entity ambiguity, a unique semantic-based mechanism is used for task entity linking and dataset entity resolution.
For online retrieval, \systemname{} utilizes a specialized hybrid query engine that combines vector search with graph-based ranking to generate highly relevant results. 
Additionally, we introduce \benchmarkname{}, a tailored benchmark suite for evaluating task-oriented dataset search systems, addressing the critical gap in standardized evaluation. Experiments on our benchmark suite show that \systemname{} significantly outperforms state-of-the-art retrieval-augmented generation frameworks in both effectiveness and efficiency, providing a robust blueprint for the next generation of dataset discovery systems. 
\end{abstract}


\maketitle

\ifdefempty{\vldbavailabilityurl}{}{
\vspace{.3cm}
\begingroup\small\noindent\raggedright\textbf{PVLDB Artifact Availability:}\\
The source code, data, and/or other artifacts have been made available at \url{\vldbavailabilityurl}.
\endgroup
}

\section{Introduction}

The search for suitable datasets is a foundational yet often arduous step in any data-driven scientific investigation. In response to this pressing need, the problem of dataset discovery has attracted growing attention from both academia and industry, leading to the development of a range of data search systems~\cite{Hulsebos2024ItTLMainSurvey, BrickleyBurgessNoy2019googledatasetsearch, Balaka2025Pneuma}. However, the dominant paradigm among these systems remains keyword-based retrieval, which implicitly assumes that users can articulate their data needs using structured metadata or keyword fields. This approach presumes that users have already translated their high-level research objectives into a precise data specification.

In contrast, recent empirical studies reveal a stark misalignment between this assumption and real-world behavior: for $79\%$ of data professionals, identifying a relevant dataset is the single most common and critical task~\cite{Hulsebos2024ItTLMainSurvey}. This underscores the importance of \emph{task-oriented} dataset search, i.e., retrieving datasets based on natural language task descriptions, which remains a long-standing and unsolved challenge~\cite{Chapman2019DatasetSurvey}.

We identify three fundamental obstacles that hinder progress in this domain:

\begin{itemize}
    \item \textbf{Ambiguous User Intent.} A substantial ``semantic gap'' exists between users’ high-level goals and the low-level keyword queries typically required by search engines. This misalignment often yields irrelevant results~\cite{Chapman2019DatasetSurvey}.
    
    \item \textbf{Task-to-Dataset Mapping and Benchmark Gaps.} Even when a system can interpret user intent, there exists no explicit mapping between scientific tasks and their associated datasets. Moreover, the lack of standardized benchmarks has hindered rigorous evaluation and comparative study of task-oriented dataset search systems.
    
    \item \textbf{Entity Ambiguity.} Datasets are inconsistently referenced across the literature, e.g., by acronym, full name, or version, 
    which leads to fragmented results and low recall~\cite{Chapman2019DatasetSurvey}.
\end{itemize}

Fortunately, recent technological advances offer new opportunities to tackle these challenges. Large Language Models (LLMs) have demonstrated impressive capabilities in semantic understanding and reasoning~\cite{zhao2023surveyLLM}, making them well-suited for interpreting complex task descriptions. In parallel, the Retrieval-Augmented Generation (RAG) paradigm~\cite{lewis2020RAG} provides a flexible mechanism for grounding LLMs in external knowledge sources.

Despite these advances, existing RAG frameworks are typically designed for general-purpose question answering, not for task-oriented dataset discovery. For instance, HippoRAG focuses on multi-hop reasoning~\cite{jimenez2024hipporag}, while Raptor’s hierarchical design may miss cross-document associations~\cite{liu2025hoprag}. Crucially, core issues such as entity ambiguity, task-dataset mapping, and relevance ranking in the context of dataset retrieval remain under-addressed. Furthermore, building and operating RAG systems at scale remains time- and resource-intensive.

To address these issues, we present \textbf{K}nowledge graph-\textbf{A}ugmented \textbf{T}ask-oriented dataset \textbf{S}earch (\systemname), a purpose-built system for task-driven dataset discovery. \systemname{} operates through a two-stage pipeline: (1) an offline component that automatically constructs a task-dataset Knowledge Graph (KG) \revised{and supports efficient incremental updates}, and (2) an online hybrid query engine that integrates LLM-powered retrieval with graph-based reasoning to surface relevant datasets.

Importantly, we also introduce a standardized \textbf{benchmark} for task-oriented dataset search, derived from scientific literature, the same type of corpus used to construct our KG. Scientific papers are a natural source for this purpose, as they routinely describe both the task and the datasets used. This implicit alignment enables us to curate realistic task–dataset pairs at scale, offering a reproducible and representative framework for evaluating dataset discovery systems.

Our key contributions are summarized as follows:

\begin{itemize}
    \item We develop a standardized benchmark suite for task-oriented dataset search, constructed from curated scientific literature. This addresses a longstanding evaluation gap and supports systematic assessment of dataset discovery methods.
    
    \item We design a novel end-to-end pipeline that automatically builds a task-dataset KG, augmented with a dedicated entity resolution module to combat dataset naming ambiguities.
    
    \item We propose a hybrid query engine that fuses dense retrieval with graph-based relevance ranking, yielding accurate and task-relevant dataset recommendations.
    
    \item Extensive experiments demonstrate that \systemname{} outperforms existing LLM-based approaches in both effectiveness and computational efficiency.
\end{itemize}

\begin{figure*}[ht]
  \centering
  \includegraphics[width=0.76\linewidth]{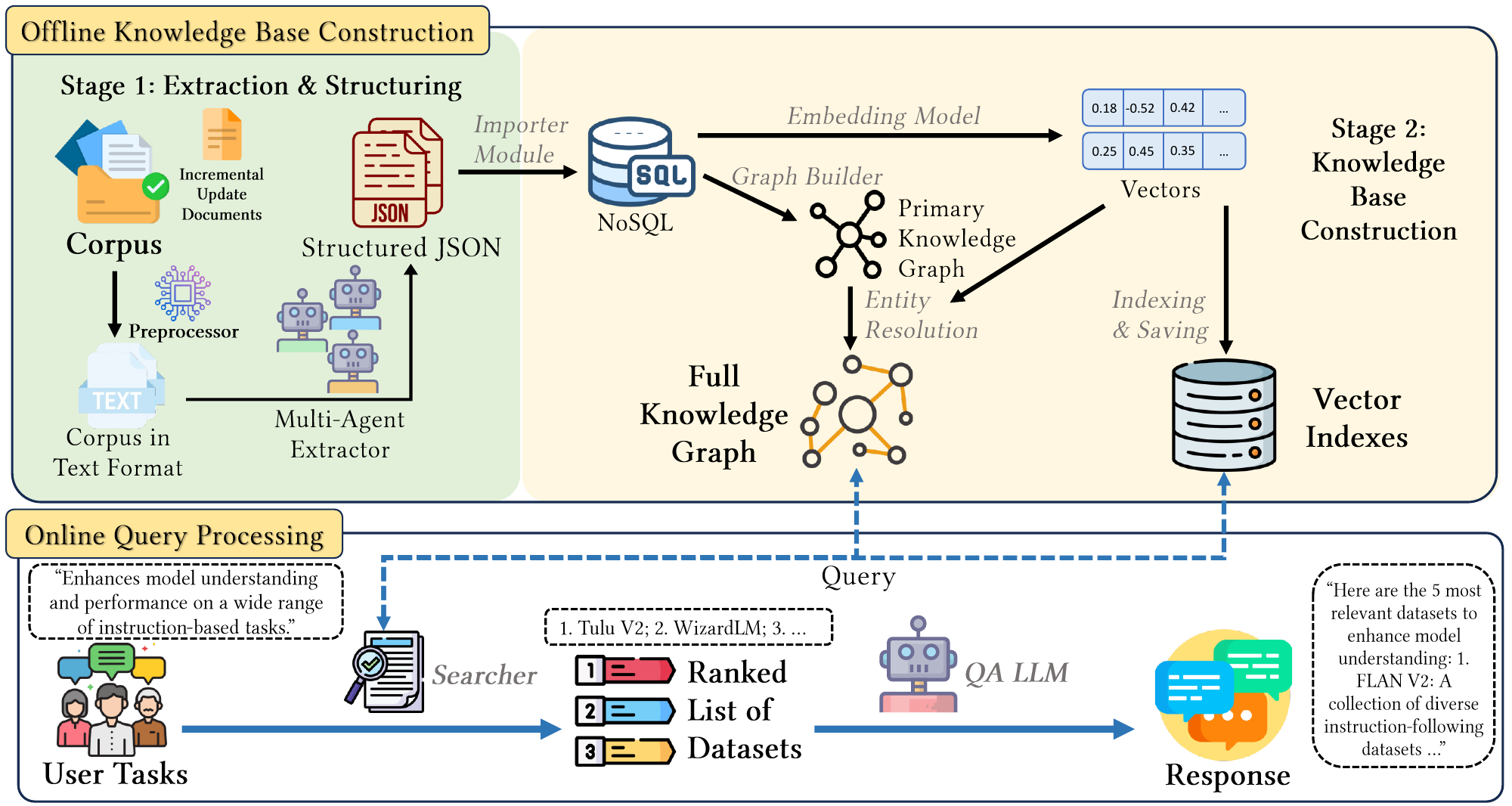}
  \caption{Overall architecture of the \systemname{} system.}
  \label{fig:system_architecture}
\end{figure*}

The remainder of this paper is structured as follows: \Cref{sec:relatedwork} surveys related work in dataset search and RAG systems. \Cref{sec:problemdef} defines the task-oriented dataset search problem and presents our benchmark. \Cref{sec:sysoverview}–\ref{sec:queryprocessing} detail the system architecture. \Cref{sec:experiments} reports experimental results, followed by discussions and conclusions in \Cref{sec:discussion,sec:conclusion}.

\section{Related Work}
\label{sec:relatedwork}

\subsection{Dataset Search Systems}

Dataset search has long been recognized as a fundamental yet challenging problem in both academia and industry. A recent survey shows that the most common goal of data search is to find the initial dataset for a given task~\cite{Hulsebos2024ItTLMainSurvey}. However, this process is challenging given the existing data quality and search methods. Existing search systems are often hampered by poor metadata quality~\cite{neum-etal-2016JDIQ}. Meanwhile, there is a semantic gap between user intent and system capabilities, including understanding user queries in the context of available datasets and matching semantically related but syntactically different datasets~\cite{PatonChenWu2023}. Our work is conducted within this extensive research context, aiming to promote the evolution of the dataset discovery paradigm from simple keyword searches to more intelligent and user-centric approaches.

\subsubsection{General and Web-scale Systems}

In the early stages of dataset search, researchers focused on building web-scale discovery tools, of which Google Dataset Search~\cite{BrickleyBurgessNoy2019googledatasetsearch} is the most representative. This system relies on data owners to publish structured metadata in accordance with open metadata standards such as \texttt{schema.org}~\cite{schema.org} and Croissant~\cite{akhtar2024croissant}. By aggregating this metadata, the system can index massive datasets. However, the fundamental limitation of this approach lies in its dependence on high-quality, manually curated metadata. When metadata is missing or of poor quality, search results are significantly degraded.

\subsubsection{Semantic-based Systems}

With the spread of the ``data lake'' concept~\cite{Bogatu2020DataLakeDatasetDiscovery}, dataset discovery faces new challenges. It has become complicated to discover relevant datasets in an environment where large amounts of heterogeneous datasets are stored in raw format with missing or inconsistent pattern information. In response to these challenges, a series of advanced semantic-based systems has emerged.
Aurum~\cite{rajan2018aurum} is a representative semantic system. It builds an enterprise KG that systematically captures syntactic relationships between data columns, enabling users to perform flexible exploratory queries. Moreover, D3L~\cite{Bogatu2020DataLakeDatasetDiscovery} proposes a distance-based framework that combines multiple types of evidence to discover related tables in data lakes. In addition, Table Union Search~\cite{Nargesian2018TableUS} enables joint searches of table data in an open data environment, improving the quality of related table discovery and search efficiency. Similarly, Auctus~\cite{castelo2021auctus} is a search engine that supports data enrichment queries, allowing users to find datasets that can be merged with a given table.

After the emergence of LLMs, systems with LLM integration have demonstrated the power of using LLMs in dataset retrieval. PNEUMA~\cite{Balaka2025Pneuma} combines LLMs with traditional information retrieval techniques to efficiently and accurately discover relevant tabular data from large collections based on natural language queries. LEDD~\cite{An2025LEDD} utilizes LLM to generate a hierarchical global directory automatically and supports semantic table search using natural language.. Furthermore, SemExplorer~\cite{semexplorer} displays a user interface designed specifically for customized semantic dataset search, highlighting the importance of human-computer interaction in solving complex discovery problems. AUTOTUS~\cite{Hu2023AutoTUS} combines the contextual understanding capabilities of LLM with its multi-stage training strategies to achieve deep representation learning of complex relationships between tables, which significantly improves the accuracy and robustness of automatic joint table search. 
This series of research studies clearly demonstrates a paradigm shift from reliance on external metadata to a deeper understanding of data content. However, this is still insufficient to fully satisfy users' needs to find datasets that serve their specific ``tasks.''

\subsubsection{Task-Driven Dataset Search}

The task-driven paradigm will be a key shift in the field of dataset search, motivated by the real-world work practices of data professionals. The recent survey of data professionals~\cite{Hulsebos2024ItTLMainSurvey} identified task-driven search as one of the four key desiderata for next-generation dataset search systems. This highlights a critical need derived directly from user practice: users are often unclear about the name of the dataset they need, but rather seek datasets suitable for accomplishing a specific task.

Recent research has begun to explore this task-driven paradigm. For instance, METAM~\cite{Galhotra2023METAM} proposes a goal-oriented framework where the ``task'' is a computable Machine Learning (ML) model. The system iteratively augments a dataset to improve a quantitative utility score, such as the F1-score. Similarly, Mileena~\cite{Huang2024FastPrivate} also defines the task as an ML model, but focuses on solving the critical challenges of search latency and data privacy by using semi-ring sketches.

While these systems are powerful and address important aspects of the problem, their definition of a ``task' relies on a formal and quantifiable function. In contrast, our work tackles a more common scenario where a user's task is expressed as an open-ended natural language description. This formulation requires a fundamentally different approach centered on a deep semantic understanding of the user's intent and its relationship to the knowledge within the unstructured text. Addressing this specific challenge of task-driven search from natural language is the primary motivation for our work.

\subsection{Knowledge Graph-based RAG}

The advent of RAG has opened up new areas for knowledge-intensive tasks. Recent research has gone beyond simple vector-based retrieval by proposing more sophisticated architectures that allow the construction of knowledge corpora for more efficient retrieval. Therefore, these approaches are applicable to the task of dataset search from unstructured scientific literature. We position our work in relation to several of these state-of-the-art frameworks, which we also employ as strong baselines in our experiments.

These advanced systems represent different strategies for structuring knowledge. GraphRAG~\cite{edge2024graphrag} constructs an entity-based KG from the source documents using LLM and identifies communities to generate abstractive summaries for answering broad questions. The HippoRAG series of works presents a neurobiologically inspired framework for enhancing the reasoning capabilities of RAG. The original HippoRAG~\cite{jimenez2024hipporag} introduced the core idea of using the Personalized PageRank on an LLM-constructed KG to enable multi-hop retrieval. Then, its successor HippoRAG 2~\cite{jimenez2025hipporag2} builds upon this foundation to address several limitations of the original entity-centric approach. Specifically, HippoRAG 2 introduces passage nodes and a more sophisticated query-to-triple matching mechanism for deeper contextualization. It also incorporates a recognition memory module to improve the retrieval accuracy and efficiency. RAPTOR~\cite{sarthi2024raptor} recursively clusters and summarizes text chunks to build a hierarchy, allowing the system to retrieve information at varying levels of abstraction for general question answering over long documents.

While these systems are powerful general-purpose frameworks, they are not specifically optimized for the unique challenges of dataset discovery from scientific literature. In contrast, our proposed \systemname{} system is an end-to-end pipeline tailored for this specific task. It incorporates specialized mechanisms, such as resolving entity ambiguity and datasets ranking, which are not the primary focus of these more general RAG architectures.

\section{Problem Definition}
\label{sec:problemdef}

This section provides a formal definition of our primary problem, \texttt{Task-Oriented Dataset Search} problem. Subsequently, we describe the construction and characteristics of our new benchmark suite, which was created to rigorously evaluate systems on this specific task.

\subsection{The Task-Oriented Dataset Search Problem}\

Let $C = \{c_1, c_2, \cdots, c_n\}$ be a corpus of unstructured scholarly documents. Given a task query $T$ in natural language from a user, the goal of a \texttt{Task-Oriented Dataset Search} system is to return a ranked list of datasets $D=\{d_1, d_2, \cdots, d_k\}$ such that each dataset $d_i$ corresponds to a unique dataset entity within the corpus $C$, and is relevant to solving the task described as $T$. The objective of the system is to maximize the relevance of the returned list of datasets $D$. The relevance of a dataset to a task query is defined according to the ground-truth annotation criteria established in our benchmark suite (see Section~\ref{sec:problem:bench}).

\subsection{Benchmark Suite}
\label{sec:problem:bench}

While several benchmarks exist for question answering (e.g., HotPotQA~\cite{yang2018hotpotqa}) or general document retrieval (e.g., BEIR~\cite{thakur2021beir}), there is a lack of established benchmarks specifically designed for evaluating task-oriented dataset search over a large corpus of scientific documents. For a rigorous and reproducible evaluation, we constructed a new benchmark suite, which we refer to as the Computer Science Task Dataset Search (\benchmarkname).

\subsubsection{Corpus Curation}

Our suite consists of two benchmarks built upon corpora of different scales to test system performance and scalability, with their detailed statistics summarized in \Cref{tbl:benchmark_stats}. The first one, $\text{\benchmarkname}_{M}$, is constructed from a corpus of $628$ computer science papers, and the second, larger benchmark $\text{\benchmarkname}_{L}$ is built from $2101$ papers. These papers were obtained using web crawlers from computer science conferences, e.g., NeurIPS.

\begin{table}[h]
  \centering
  \caption{Statistics of \benchmarkname{} benchmark suite.}
  \label{tbl:benchmark_stats}
  \begin{tabular}{lrr}
    \toprule
    \textbf{Statistic} & \textbf{\benchmarkname{}$_M$} & \textbf{\benchmarkname{}$_L$} \\
    \midrule
    \# Papers in Corpus & $628$ & $2,101$ \\
    \# Task Queries & $47$ & $204$ \\
    \# Datasets & $\sim 1,779$ & $\sim 7,525$ \\
    \# Unique Datasets & $\sim 1,217$ & $\sim 4,180$ \\
    Year Span & $2023$ & $2024$ \\
    \bottomrule
  \end{tabular}
\end{table}

\subsubsection{Query Generation}

The queries are generated as natural language descriptions of a task derived from real papers by an LLM and manually verified. $47$ queries are selected for $\text{\benchmarkname}_{M}$, and $204$ are selected for $\text{\benchmarkname}_{L}$. To ensure the evaluation tests a system's generalization ability rather than its ability to simply retrieve the source document, the paper from which the query and ground-truth answer were derived is held out from the search corpus. Below is a representative example from our benchmark:

\begin{tcolorbox}[
  colback=black!5,
  colframe=black!75,
  title=\textbf{Benchmark Query Example},
  fonttitle=\small\bfseries,
  arc=2mm,
  boxrule=0.5pt
]
\small
\textbf{Task Query:} \textit{``Assess HSIVI-SM's ability to accelerate diffusion model sampling while maintaining sample quality, measured via Frchet Inception Distance (FID).''}
\vspace{1mm}

\textbf{Ground-Truth Dataset:} \textit{``CIFAR-10''}, which contains $60,000$ $32 \times 32$ color images in $10$ different classes~\cite{cifar10}.
\end{tcolorbox}

\subsubsection{Ground-Truth Annotation}

For each query, the standard answer is set as the corresponding dataset actually used in the paper. To avoid ``false negatives,'' other reasonable answers besides the original answer are also considered correct answers.

Formally, a dataset $d_i$ is considered \textit{relevant} to a task query $T$ if it satisfies one of the following conditions during ground-truth annotation:

\begin{itemize}
    \item It is \textbf{identical} to the dataset used in the source paper for $T$.
    \item It is a \textbf{variant or alias} of the dataset (e.g., MIMIC-III vs. Medical Information Mart for Intensive Care III).
    \item It is a \textbf{functionally equivalent substitute}, typically from the same dataset family or with the same task objective (e.g., LAION-5B vs. LAION-400M for large-scale image-text pretraining).
\end{itemize}

These criteria and the corresponding annotations are manually validated and used to determine the binary \textit{relevance label} during evaluation. Systems are scored based on how well they retrieve such relevant datasets for each query.

\section{System Overview}
\label{sec:sysoverview}

In this section, we introduce the design of our proposed \systemname{} in detail. Firstly, we present the overall architecture and the function of each component. Then, we formally define the data model of our KG, which is the core data structure managed by the system.

\subsection{Overall Architecture}

As illustrated in Figure \ref{fig:system_architecture}, the architecture of \systemname{} is divided into two major phases: an offline phase for knowledge base construction and an online phase for query processing.

\subsubsection{Offline Knowledge Base Construction}

The offline phase begins with \texttt{Stage 1: Extraction \& Structuring}. In this process, the input papers are transformed into a structured format for the following knowledge base construction. The \texttt{Preprocessor} module normalizes heterogeneous source documents (mostly in PDF format) into the plain text representation. Then, the downstream \texttt{Extractor} module uses an LLM to extract task and dataset information into JSON format.

Structured information is then passed to \texttt{Stage 2: Knowledge Base Construction}. It involves two interconnected workflows after structured information is imported into the database. The vector workflow generates dense vector representations for tasks and datasets, and stores them in the \texttt{Vector Indexes}. Meanwhile, in the graph workflow, the \texttt{GraphBuilder} module populates the initial KG. Subsequently, the \texttt{Entity Resolution} module further refines the KG with the previously generated vector embeddings.

\subsubsection{Online Query Processing}

The online phase is designed for real-time user interaction. When a user submits a \texttt{Natural Language Task Query}, the \texttt{Hybrid Searcher} interacts with the offline knowledge base. It queries the \texttt{Vector Indexes} for semantic similarity and the \texttt{KG} for relational task-task and task-dataset information. Combining these signals, the searcher produces a \texttt{Ranked List of Datasets} that are most relevant to the user's task. This list is then passed to a downstream \texttt{QA LLM}, which can re-rank the results for better precision and generate a final natural language answer for the user.

\subsection{Task-Dataset KG}

The core data structure managed by \systemname{} is the \textit{Task-Dataset KG}. This graph is formally defined as $G = (V,E)$, where $V$ represents the set of nodes (vertices) and $E$ represents the set of undirected edges.

\begin{table}[h]
\centering
\caption{Notations and meanings.}
\begin{tabular}{c|p{6cm}}
\toprule
\textbf{Notation} & \textbf{Meaning} \\
\midrule
\multirow{3}{*}{\( G = (V, E, W) \)} & Undirected, weighted KG with vertex set \( V = C \cup D \cup T \), edge set \( E = E_{C,D} \cup E_{D,T} \cup E_{T,T} \), and weight function \( W \). \\
\hline
\( C \) & Set of document nodes. \\
\hline
\( D \) & Set of dataset nodes. \\
\hline
\( T \) & Set of task nodes. \\
\hline
\( E_{C,D} \) & Set of edges \( (c, d) \), where \( c \in C \), \( d \in D \). \\
\hline
\( E_{D,T} \) & Set of edges \( (d, t) \), where \( d \in D \), \( t \in T \). \\
\hline
\( E_{T,T} \) & Set of edges \( (t_1, t_2) \), where \( t_1, t_2 \in T \). \\
\bottomrule
\end{tabular}
\end{table}
The set of vertices $V$ consists of three distinct types of nodes:

\begin{itemize}
    \item \textbf{Document Node}: Represents a single document (paper) from the input. Each node stores metadata of the document, such as title and source file path.
    \item \textbf{Dataset Node}: Represents a dataset entity used in the document. Each node includes attributes such as the dataset's name and brief description. 
    \item \textbf{Task Node}: Represents a specific task for which a dataset is used, as described within a document. Each node contains attributes such as a textual description of the task.
\end{itemize}

The set of edges $E$ captures the relationships between these nodes:

\begin{itemize}
    \item \textbf{Document-dataset Edge}: An edge $(c, d)$ where $c \in C$, $d \in D$ signifies that document $c$ includes dataset $d$.
    \item \textbf{Dataset-task Edge}: An edge $(d, t)$ where $d \in D$, $t \in T$ signifies that dataset $d$ is used for the purpose of task $t$.
    \item \textbf{Task-task Edge}: An edge $(t_1 ,t_2)$ where $t_1, t_2 \in T$ with weight $w$ signifies that task $t_1$ and $t_2$ are semantically similar with cosine similarity $w$.
\end{itemize}

\section{Knowledge Base Construction}
\label{sec:knowledgebase}

\begin{figure}[ht]
  \centering
  \includegraphics[width=0.66\linewidth]{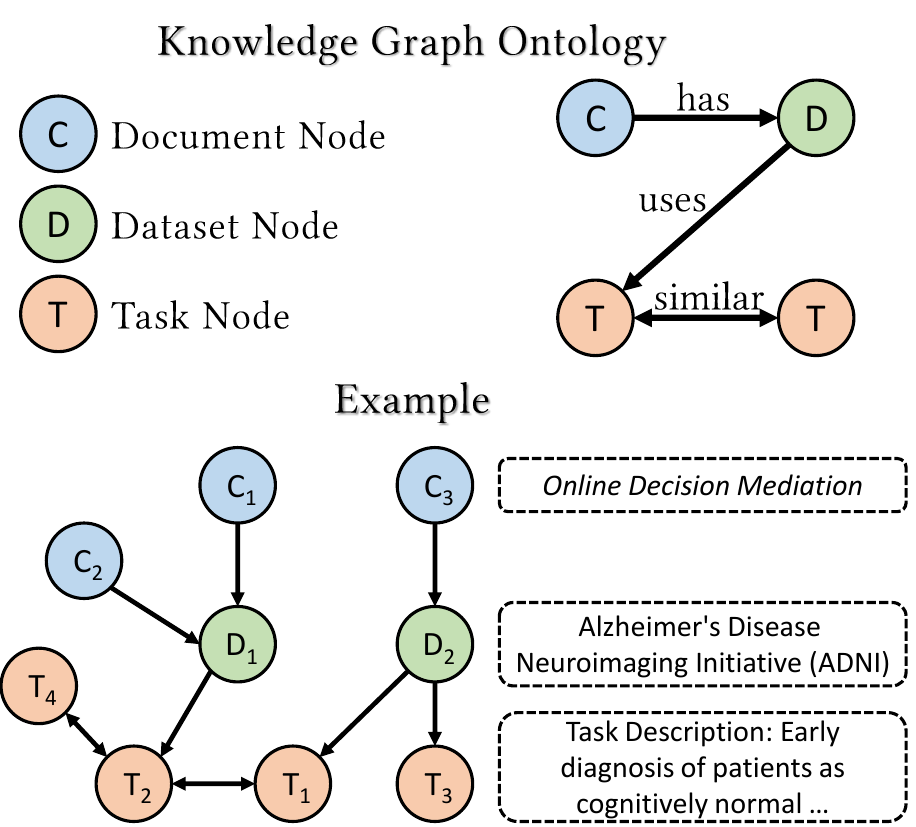} 
  \caption{Ontology and an example of the task-dataset KG.}
  \label{fig:kg}
\end{figure}

In this section, we will introduce the details of the offline pipeline for constructing the task-dataset knowledge base. 
While contemporary graph-based RAG systems also construct KGs from text, their methodology is often tailored for general-purpose summarization. However, for precise task-oriented dataset search, this approach can be suboptimal, as critical details such as the context of a task may be lost. Furthermore, these general frameworks often lack the dedicated mechanisms required to address the entity ambiguity problem. 
To overcome these limitations, our pipeline is specifically designed for the precise extraction, curation, and querying of dataset and task entities. The process starts with raw documents and produces a high-quality, queryable knowledge base within a series of steps. 
\revised{Furthermore, to ensure the long-term viability and scalability of the knowledge base, we designed the system to support efficient incremental updates, allowing it to remain up-to-date without costly periodic rebuilds.} 
Each step will be further discussed in the following subsections. The ontology and an example of the task-dataset KG are demonstrated in Figure~\ref{fig:kg}.

\subsection{Document Normalization and Caching}

The first step in the pipeline is to collect and normalize the source documents. The primary goal of this step is to convert a corpus of documents in heterogeneous formats into plain text. The main challenge lies in handling the complexity of PDF format papers to ensure that the LLM can accurately extract information in the subsequent step.

In our current implementation, we utilize the pypdf~\cite{pypdf} library to employ direct text extraction, which is highly efficient. Since our target entities - datasets and their corresponding tasks - primarily appear in the narrative prose of the papers, non-textual elements such as figures and tables play a less critical role in our extraction strategy. Therefore, this method proves sufficient and effective for reliably extracting our target entities. To address more complex cases, such as when many texts are presented in the form of images or complex multi-column layouts, we could consider Optical Character Recognition (OCR) or intelligent document parsing services (e.g., LlamaParse~\cite{llamaparse}), which is computationally expensive. This step is completed in a completely separate module. Thus, the specific method can be replaced depending on the condition of the source documents to increase efficiency and save costs.

To optimize performance, especially when dealing with large, static corpora, we implement a caching strategy. Before processing each file, the system computes a unique hash value of the file's content as its ``fingerprint''. On subsequent runs, if a file with an identical fingerprint is encountered, the content will be retrieved directly from cache. This prevents costs associated with duplicate computing. We utilize the same caching approach in the subsequent information extraction step.

\subsection{Information Extraction}

Once the documents are normalized into plain text, the \texttt{Extractor} module is responsible for identifying and extracting structured information about datasets and their associated tasks. Some contemporary methods, such as GraphRAG~\cite{edge2024graphrag}, perform a one-shot, general-purpose entity extraction. Instead, we design a collaborative multi-agent framework specifically for our goal of task-oriented dataset discovery. The framework decomposes the complex extraction challenge into a sequence of sub-tasks, each handled by a specialized agent. This design not only improves the accuracy and relevance of the extracted information but also enhances efficiency by avoiding deep processing of irrelevant documents.

\subsubsection{The Preliminary Filtering Agent}

The first agent in our framework acts as an efficient gatekeeper. Its primary responsibility is to rapidly identify documents that are likely to contain dataset mentions to avoid expensive LLM calls on irrelevant content. For each document, this agent queries the LLM with a concise instruction template designed to obtain a simple Boolean response of dataset relevance. Only documents identified to be relevant are passed to the subsequent agent.

\subsubsection{The Analyst Agent}

The Analyst Agent specializes in deep content analysis. It utilizes a detailed instruction template to guide the LLM's careful reading of the text to help identify all mentions of datasets and their associated tasks. To ensure machine readability, this agent is constrained to generate its output in a predefined structured JSON format. This format consists of a list of objects, where each object represents a unique dataset-task relationship and contains relevant information. Table \ref{tbl:extraction_example} presents a concrete example of an input text snippet in a paper~\cite{jarrett2022online} and the corresponding structured JSON object generated by the agent.

\subsubsection{The Enrichment Agent}

Finally, the structured JSON object is passed to the Enrichment Agent. This agent's sole purpose is to enhance the semantic metadata of the extracted entities. For each task description, the agent uses a separate instruction template to generate a list of relevant keywords that summarize the core concepts of the task. These keywords will help graph curation in the subsequent stage.

\begin{table*}[ht]
  \centering
  \caption{Example of the LLM-based information extraction process.}
  \label{tbl:extraction_example}
  \begin{tabularx}{\textwidth}{l X}
    \toprule
    \textbf{Component} & \textbf{Content} \\
    \midrule
    \textbf{Input Text Snippet} & In Alzheimers, the task is to perform early diagnosis of patients in the Alzheimer’s Disease Neuroimaging Initiative study as cognitively normal, mildly impaired, or at risk of dementia. \\
    \midrule
    \textbf{Generated JSON Output} & 
    \{"dataset\_name": "Alzheimer's Disease Neuroimaging Initiative (ADNI)", "task\_description": "early diagnosis of patients as cognitively normal, mildly impaired, or at risk of dementia.", "task\_keywords": ["early diagnosis", "dementia", "patient assessment"]\}
    \\
    \bottomrule
  \end{tabularx}
\end{table*}

\subsection{Semantic Representation Learning}

After extracting structured textual information, the next step is to convert this text into meaningful vector representations. This process maps textual descriptions into a high-dimensional vector space. These vector embeddings are fundamental to our system, for both quantitative analysis of similarity in graph building and the semantic search capabilities in online querying.

In our system, we generate embeddings for two key types of entities: tasks and datasets. For each task node, we embed its task description. For each dataset node, we concatenate its title and description to create a comprehensive representation. In our implementation, we utilize OpenAI's text-embedding-3-small model~\cite{openai2024embedding} via its API. We selected this model for its compelling balance of high embedding quality, cost-effectiveness, and ease of integration.

The generated embeddings, $\{e_{task}\}$ and $\{e_{dataset}\}$, are then persisted and indexed for efficient retrieval. Specifically, we utilize the Faiss library~\cite{douze2024faiss} to build an index from the dense vectors. Faiss enables highly efficient approximate nearest neighbor search. It allows our system to find the most semantically similar items from a large collection in real-time. This indexing step ensures that both the online semantic search and the offline graph-building processes can perform similarity lookup with low latency.

\subsection{Entity Resolution and Linking}

The initial KG constructed from the extraction stage, while comprehensive, still contains ambiguity and redundancy inherent in natural language, especially for the task and dataset nodes:
\begin{itemize}
\item Semantically identical tasks may be described with different phrasing, such as ``image classification'' and ``classifying images''.
\item The same dataset may be referred to by different names or acronyms, such as ``COCO'' and ``MS-COCO''.
\end{itemize}
To enhance the quality and reliability of the knowledge base, we perform a graph curation stage consisting of two main processes: task linking and dataset resolution.

\subsubsection{Task Linking}

To address the ambiguity of task descriptions, our system identifies and connects task nodes that are semantically similar but are described using different phrasing. To achieve this goal, the \texttt{TaskMerger} module uses the vector embeddings $\{e_{task}\}$. This process involves using Faiss to find pairs of tasks with high cosine similarity of the vectors. If the similarity score between two tasks, $t_1$ and $t_2$, exceeds a predefined threshold $\theta_d$, or the keyword overlap exceeds a threshold $\theta_k$, we consider them semantically closely related.

After identifying such task-task pairs, a weighted edge is added to the KG between the corresponding task nodes. The weight of this edge is set to the calculated similarity score, representing the strength of the relationship. These similarity edges are crucial for the KG-based task expansion in the online phase. When querying, the system is able to broaden the search to a wider set of relevant tasks through the edges.

\subsubsection{Dataset Resolution}

The dataset resolution process tackles the problem of entity redundancy, where the same dataset is referred to by different names or acronyms. An all-pairs comparison of dataset nodes would be $\mathcal{O}(n^2)$, which is computationally prohibitive. To address this problem, we designed an efficient and effective three-stage approach.

The first stage focuses on efficiently generating candidate pairs for potential merging. Instead of a brute-force comparison, we use the Faiss index. For each dataset, the system performs an approximate nearest neighbor search to retrieve a small set of the most semantically similar datasets. This process reduces the search space from all possible pairs to a manageable set of high-potential candidates. At the beginning, each dataset is stored as a separate node in a disjoint-set data structure.

In the second stage, the system performs the initial matching by normalizing the titles. Specifically, case, spaces, etc., are standardized. In addition, a dictionary is maintained to record the aliases of the dataset. After these processes, datasets with the same normalized title are merged into the disjoint set.

The above approach can partially reduce the problem of the same dataset with different names, such as ``Image Net'' and ``imagenet''. For more complex situations, an LLM works as an expert judge to perform a rigorous verification process in the third stage. The prompt provides the names and descriptions of both datasets and instructs it to return a binary ``True'' or ``False'' answer on whether they refer to the exact same entity. If the LLM returns a positive judgment, the system proceeds to merge the two nodes in the KG. Meanwhile, regardless of the judgment, the result is stored in cache to avoid repeated LLM calls.

This hybrid three-stage approach combines the scalability of vector search with the reasoning capabilities of the LLM. Therefore, our system provides a robust and efficient solution to the dataset entity resolution problem.

\revised{
\subsection{Incremental Knowledge Base Updates}
\label{sec:incremental_updates}

To address the limitation of a static knowledge base and ensure the system remains up-to-date, we designed an efficient incremental update pipeline. This mechanism allows \systemname{} to integrate the newly extracted knowledge into the existing KG and vector indices without requiring a full, costly reconstruction of the entire knowledge base.

The pipeline first identifies new documents by maintaining a manifest of all previously processed files. Each file is identified by its unique content-based fingerprint, which is generated by extracting text from the file and applying the SHA256 algorithm. When new documents are added to the corpus, the system computes their fingerprints and compares them against the manifest to create a queue of only the new files. Each new document undergoes the preprocessing and multi-agent extraction pipeline, and the extracted structured data is then ingested into the database. This process is inherently incremental since new entries are simply added to their respective tables without affecting existing data.

To update the semantic search indices, vector embeddings are generated for these new entries and are appended to the existing Faiss indices. Afterwards, new nodes, including documents, datasets, tasks, and their basic relationships, are added to the existing KG. Subsequently, the entity linking and resolution modules will operate in an incremental mode. For each new task and dataset node, the system computes its similarity against the entire index to establish new ``similar\_task'' edges and identify potential dataset merges. This approach ensures that new entities are correctly integrated into the existing graph structure while preserving all previously computed relationships. This entire pipeline ensures that \systemname{} is not merely a static repository, but a dynamic system that evolves its knowledge over time.

Furthermore, to utilize broader dataset information beyond scientific literature, the incremental update feature also incorporates the ability to integrate datasets lacking explicit task descriptions. By embedding the new description and identifying the most similar dataset within the existing knowledge graph, \systemname{} can automatically link the new entry to relevant tasks, thus enriching the knowledge base from a wider range of sources.
}

\section{Query Processing}
\label{sec:queryprocessing}

\begin{figure}[ht]
  \centering
  \includegraphics[width=0.77\linewidth]{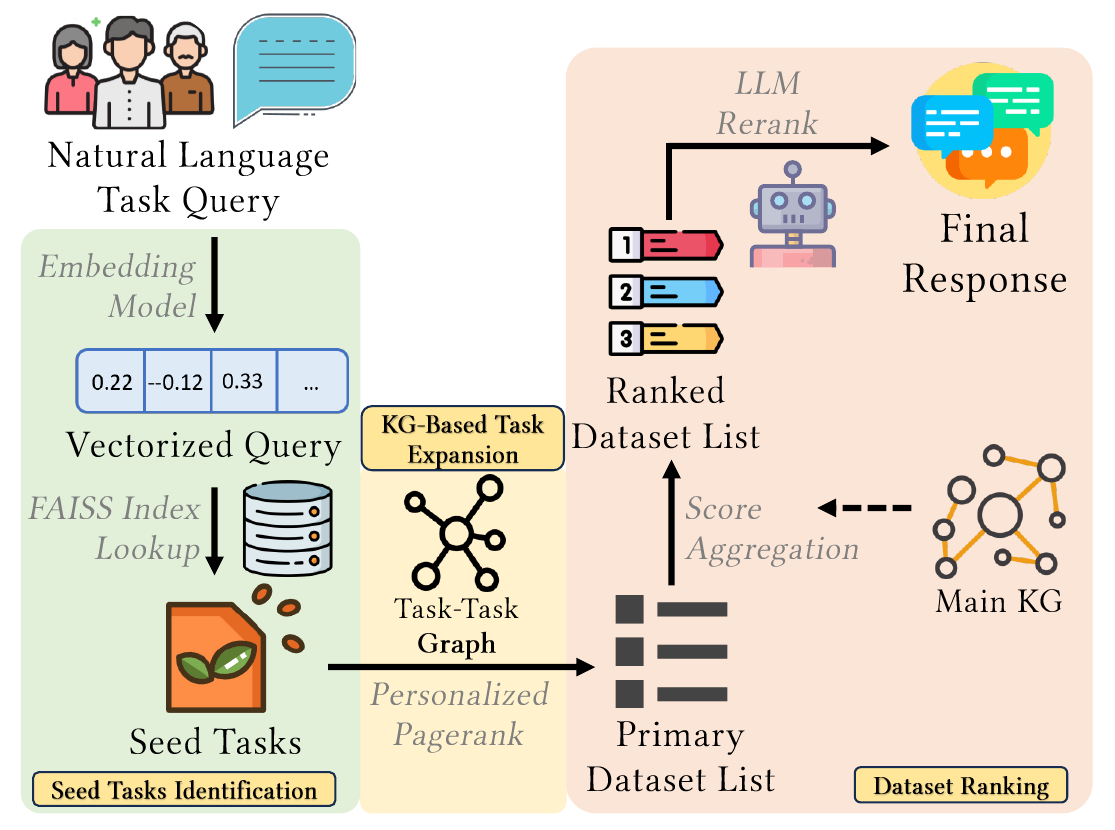} 
  \caption{The query processing pipeline of \systemname{} system.}
  \label{fig:query_flowchart}
\end{figure}

Figure~\ref{fig:query_flowchart} illustrates the entire workflow of the online query processing pipeline, which is triggered when a user submits a natural language task query. Our approach is designed as a multi-stage funnel that progressively refines the search space from broad semantic matching to fine-grained graph-based ranking. 

\subsection{Seed Task Identification}

The online query processing pipeline begins with the seed task identification stage. The primary objective of this stage is to efficiently bridge the semantic gap between the user's natural language query and the structured entities within our KG. It achieves this by identifying an initial set of ``seed'' tasks that are most semantically relevant to the user's query.

This stage involves two main steps. Firstly, the user's task query $T$ is transformed into a high-dimensional vector representation $\textbf{t}$. This is accomplished using the same pre-trained embedding model as used during the offline knowledge base construction. This ensures that both the query and the knowledge base are in the same semantic space for meaningful comparisons. Next, the query vector $\textbf{t}$ is used to perform an Approximate Nearest Neighbor (ANN) search against the pre-built FAISS index of task embeddings. This search efficiently retrieves the top-k task nodes from the KG, which we refer to as the ``seed tasks'' ($T_{seed}$). $T_{seed}$ serves as the crucial starting point for the subsequent KG-based task expansion to discover a wider range of relevant information.

\subsection{KG-based Task Expansion}

The initial set of seed tasks, $T_{seed}$, while semantically close to the query, may not be exhaustive. The KG-based task expansion stage uses the relational information in our task-task similarity graph to discover a wider set of relevant tasks. Our approach employs the Personalized PageRank (PPR) algorithm~\cite{personalizedpagerank} for this expansion. PPR allows us to compute node relevance relative to a specific set of ``source'' nodes. In our context, the algorithm is executed on the task-task similarity graph. The previously identified seed tasks $T_{seed}$ are used as the personalization vector for PPR, which means the ``random walk'' process is biased to teleport back to these seed nodes. Thus, the relevance scores from the initial candidates are effectively diffused throughout the graph. The parameter $\alpha$ balances the exploration of the graph structure against the influence of the initial seed set.

The output of this stage is a globally-aware relevance score for every task node in the KG. This score reflects both the task's direct similarity and its interconnectedness within the broader network of related tasks. The enriched set of task scores forms the input for the final dataset aggregation and ranking stage.

\subsection{Candidate Dataset Aggregation and Ranking}

In the final stage, the system aggregates the task-level relevance scores to produce a final ranked list of datasets. This process involves two key steps: an initial score aggregation based on the graph structure, followed by a final re-ranking using an LLM to enhance precision.

To aggregate candidate datasets, the system identifies the set of all dataset nodes that are connected to the high-scoring task nodes in our main KG. To calculate a relevance score for each unique dataset, we employ an aggregation strategy. The score of a dataset is defined as the maximum relevance score among all the task nodes that are linked to it:

$$Score(d) = \max_{t \in \text{LinkedTasks}(d)} \{ Score(t) \}$$

Based on these aggregated scores, all candidate datasets are then sorted in descending order to produce an initial ranked list.

To further enhance the precision of the top results, we introduce a final LLM-based re-ranking step. The system selects the top $ k_{rerank}$ datasets from the initial list. For each of these candidates, it retrieves their detailed descriptions from the knowledge base. This set of candidate descriptions, along with the original user query, is then formatted into a specialized instruction. The LLM is instructed to act as an expert evaluator and re-rank these few candidates based on their direct relevance to the user's task. This re-ranking step utilizes the superior understanding and reasoning capabilities of the LLM to provide a highly refined and precise final output to the user.

\section{Experimental Evaluation}
\label{sec:experiments}

\subsection{Experimental Setup}

\subsubsection{Datasets and Benchmarks}

Our experiments are conducted on our benchmark suite \benchmarkname. To ensure a fair and rigorous evaluation, we manually compare the ground truth with the output for each query to reduce false negatives. In addition to the original dataset, any other dataset that belongs to the same series (e.g., different scales of a dataset) and serves the same purpose was also marked as a correct answer.

\subsubsection{Baseline Systems}

To evaluate the effectiveness of \systemname, we compare it against a suite of strong baseline methods. These include an established dense retrieval approach as well as several state-of-the-art RAG systems designed for complex information retrieval and question answering:

\begin{itemize}[leftmargin=*]
    \item \textbf{VanillaRAG}: As an established semantic baseline, we utilize the Dense Passage Retrieval (DPR)~\cite{karpukhin-etal-2020-dense}. This method first splits the source documents into smaller passages. Then, it uses a pre-trained bi-encoder model to generate a dense vector representation for each passage and for the input query. Retrieval is performed by conducting a maximum inner-product search to find the most semantically relevant passages.
    \item \textbf{GraphRAG}: GraphRAG~\cite{edge2024graphrag} uses LLMs to build a KG from source text, where nodes represent extracted entities and edges represent their relationships. Retrieval is performed by interpreting the user's query to identify relevant communities or paths within the graph. We adapt it to our task by feeding our pre-processed text corpus into its graph construction engine.
    \item \textbf{HippoRAG 2}: HippoRAG 2~\cite{jimenez2025hipporag2} uses Open Information Extraction (OpenIE) to identify facts and entities within documents, and links them to the original text passages in a graph structure. The retrieval process applies the Personalized PageRank algorithm on the graph to find the most relevant information.
    \item \textbf{Raptor}: Raptor~\cite{sarthi2024raptor} recursively clusters text chunks and generates summaries for each cluster to build a tree-structured index. This hierarchical structure allows it to match a query against text at multiple levels of abstraction. In our experiments, the Raptor implementation is from DIGIMON~\cite{zhou2025depth}.
\end{itemize}

\begin{table*}[ht]
  \centering
  \caption{Detailed Effectiveness Comparison on $\text{\benchmarkname}_{M}$ and $\text{\benchmarkname}_{L}$. For each metric, the \textbf{best} result is in bold and the \underline{second-best} is underlined.}
  \label{tbl:detailed_results}
  \begin{tabular}{l cc cc cc cc cc cc}
    \toprule
    \textbf{System} & \multicolumn{2}{c}{\textbf{Hit Rate@1}} & \multicolumn{2}{c}{\textbf{Hit Rate@3}} & \multicolumn{2}{c}{\textbf{Hit Rate@5}} & \multicolumn{2}{c}{\textbf{Hit Rate@10}} & \multicolumn{2}{c}{\textbf{EM (Top-1)}} & \multicolumn{2}{c}{\textbf{F1 (Top-1)}} \\
    
    \cmidrule(lr){2-3} \cmidrule(lr){4-5} \cmidrule(lr){6-7} \cmidrule(lr){8-9} \cmidrule(lr){10-11} \cmidrule(lr){12-13}
    & $C_M$ & $C_L$ & $C_M$ & $C_L$ & $C_M$ & $C_L$ & $C_M$ & $C_L$ & $C_M$ & $C_L$ & $C_M$ & $C_L$ \\
    \midrule
    \textbf{\systemname{} (ours)} & \textbf{0.447} & \textbf{0.446} & \textbf{0.638} & \textbf{0.672} & \textbf{0.723} & \textbf{0.730} & \textbf{0.787} & \textbf{0.804} & \underline{0.208} & \textbf{0.206} & \textbf{0.350} & \textbf{0.312} \\
    \midrule
    GraphRAG & \underline{0.340} & \underline{0.309} & \underline{0.468} & 0.475 & \underline{0.596} & 0.534 & 0.617 & 0.601 & 0.064 & 0.034 & \underline{0.260} & 0.192 \\
    HippoRAG 2 & 0.298 & 0.304 & 0.426 & \underline{0.515} & 0.447 & \underline{0.608} & 0.468 & \underline{0.662} & 0.170 & 0.093 & 0.246 & 0.192 \\
    Raptor & \underline{0.340} & 0.284 & 0.404 & 0.431 & 0.511 & 0.520 & \underline{0.638 }& 0.583 & \textbf{0.213} & \underline{0.113} & 0.256 & \underline{0.205} \\
    \midrule
    VanillaRAG & 0.255 & 0.221 & 0.404 & 0.358 & 0.426 & 0.422 & 0.468 & 0.471 & 0.149 & 0.064 & 0.206& 0.149 \\
    \bottomrule
  \end{tabular}
\end{table*}
\subsubsection{Evaluation Metrics}

We evaluate the performance of all systems from two perspectives, i.e., effectiveness and efficiency.\\
\textbf{Effectiveness Metrics}
\begin{itemize}[leftmargin=*]
\item \textbf{Hit Rate@$k$}: The proportion of queries for which the correct dataset is found within the top-$k$ ranked results. We report for $k$ in {1, 3, 5, 10}.
\item \textbf{Exact Match (EM)}: For the top-1 result, this metric is 1 if the canonical name of the retrieved dataset exactly matches a ground-truth answer, and 0 otherwise.
\item \textbf{F1-Score}: Also for the top-1 result, we compute the token-level F1-score between the predicted dataset name and the ground-truth name to account for partial matches.
\end{itemize}
\textbf{Efficiency and Cost Metrics}
\begin{itemize}[leftmargin=*]
    \item \textbf{Offline Build Time}: The total wall-clock time required to execute the entire offline knowledge base construction pipeline.
    \item \textbf{Storage Footprint}: The total disk space occupied by all generated artifacts, possibly including the KG, vector indexes, and any caches.
    \item \textbf{Build Token Cost}: The total number of input and output tokens consumed by all LLM API calls during the offline build phase.
    \item \textbf{Per-Query Time Cost}: The average wall-clock time for a single query during the online phase.
    \item \textbf{Per-Query Token Cost}: The average number of tokens consumed by LLM API calls for a query during the online phase.
\end{itemize}

\subsubsection{Implementation Details}

For all experiments, our \systemname{} system was configured to use OpenAI's \texttt{text-embedding-3-small}~\cite{openai2024embedding} as the embedding model. The LLM used for all extraction and judgment tasks was \texttt{gpt-4o-mini}~\cite{openai2024gpt4omini}. Key hyperparameters for our \systemname{} searcher are:
\begin{itemize}[leftmargin=*]
    \item \textbf{Initial Candidate Size ($k$)}: The number of initial candidate tasks retrieved from the FAISS index for each query is set to $k=2$. This parameter controls the breadth of the initial semantic search.
    \item \textbf{PageRank Alpha ($\alpha$)}: The alpha parameter for the Personalized PageRank algorithm, which balances the influence of the initial query vector and the graph structure, is set to $\alpha=0.85$.
\end{itemize}

All experiments were conducted on a server equipped with dual Intel Xeon Gold 6330 CPUs, an NVIDIA RTX A5000 GPU with $24$GB of VRAM, and $1$TB of RAM. The system was running on a Linux distribution with CUDA version $12.8$.

\subsection{End-to-End Effectiveness}
\begin{figure}[ht]
  \centering
  \includegraphics[width=0.82\linewidth]{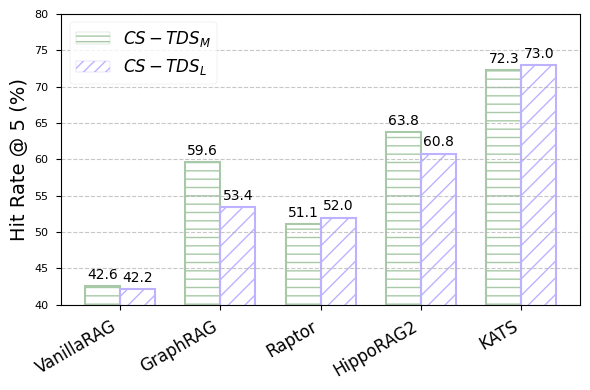}
  \caption{End-to-end effectiveness comparison in terms of Hit Rate@5 on $\text{\benchmarkname}_{M}$ and $\text{\benchmarkname}_{L}$.}
  \label{fig:main_results_chart}
\end{figure}

In this section, we present the end-to-end effectiveness evaluation of our proposed \systemname{} system against the baseline methods. Figure~\ref{fig:main_results_chart} visualizes the primary comparison in terms of Hit Rate@5. As the results clearly indicate, \systemname{} consistently and significantly outperforms all baselines across both $\text{\benchmarkname}_{M}$ and $\text{\benchmarkname}_{L}$.

On the larger dataset $\text{\benchmarkname}_{L}$, for instance, \systemname{} achieves a Hit Rate@5 of $73.0\%$. This represents a substantial improvement of nearly $20\%$ percentage points over the strongest baseline, HippoRAG 2 ($60.8\%$). The performance gap is more significant when compared to the VanillaRAG baseline, which highlights the effectiveness of our KG-based hybrid search mechanism. We observe a similar trend on the $\text{\benchmarkname}_{M}$ dataset, confirming the robustness of our approach across different corpus scales.

The significant performance advantage of \systemname{} can be attributed to several key design choices. Firstly, unlike general-purpose frameworks, our pipeline constructs a highly specialized Task-Dataset KG, focusing only on the core entities and relationships relevant to our goal. Secondly, our novel LLM-based dataset entity resolution process effectively cleans the duplicated content in the graph by merging nodes with different names but identical semantics. This curation step is critical for consolidating knowledge and improving recall. Most importantly, our hybrid search mechanism performs a sophisticated task expansion using a PPR algorithm. This allows the system to understand user queries in a broader context and move beyond initial keyword or vector matches to discover datasets associated with a wider range of semantically related tasks. This directly contributes to its superior performance over more direct retrieval methods.

The superiority of our approach is not limited to the Hit Rate@5 metric alone. For a more comprehensive breakdown across all effectiveness metrics, Table~\ref{tbl:detailed_results} presents a detailed comparison. This demonstrates the ability of \systemname{} to accurately identify the canonical name of the correct dataset. Even if not entirely precise, the search results are highly accurate in content. In summary, the combination of high recall and high precision provides strong evidence on multiple aspects of the end-to-end effectiveness of the \systemname{} framework.

\subsection{Efficiency and Cost Analysis}

\begin{figure}[ht]
  \centering
  \includegraphics[width=0.85\linewidth]{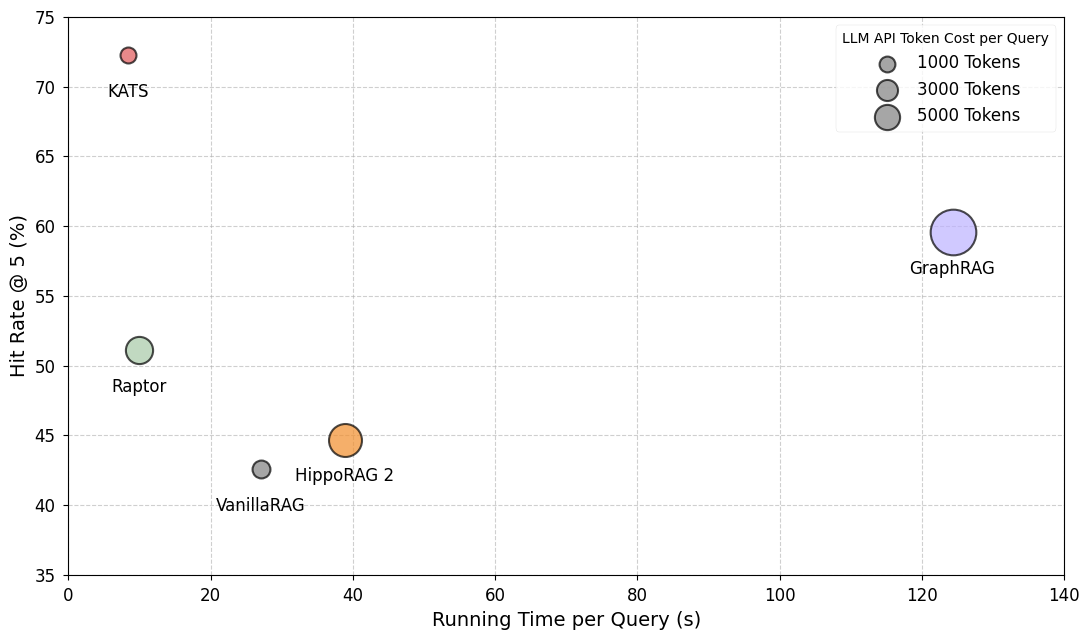} 
  \caption{Effectiveness vs. efficiency trade-offs in \systemname.}
  \label{fig:trade-off}
\end{figure}

Beyond effectiveness, the practical usability of a dataset search system depends on its efficiency and operational cost. In this section, we evaluate \systemname{} and the baseline methods across three key dimensions: time cost, API cost (token consumption), and storage cost. Our analysis is conducted on the $\text{\benchmarkname}_{M}$ benchmark. A holistic view of the trade-offs across effectiveness and efficiency is presented in Figure~\ref{fig:trade-off}.

\begin{table}[ht]
  \centering
  \caption{Comparison of time, API, and storage costs during offline building (B.) and online retrieving (R.) on $\text{\benchmarkname}_{M}$.}
  \label{tbl:overall_cost}
  \begin{tabular}{l c c c c c}
    \toprule
    \multirow{2}{*}{\textbf{System}} 
      & \multicolumn{2}{c}{\textbf{Time}} 
      & \multicolumn{2}{c}{\textbf{API}} 
      & \multirow{2}{*}{\shortstack{\textbf{Storage}\\\textbf{(MB)}}} \\
    \cmidrule(lr){2-3} \cmidrule(lr){4-5}
      & B. (h) & R. (s)
      & B. (M) & R.
      & \\
    \midrule
    \textbf{\systemname{} (ours)} & $5.2$ & $8.4$ & $21$ & $1,041$ & $255$\\
    \midrule
    GraphRAG & $99.4$ & $124.3$ & $151$ & $20,011$ & $8,228$\\
    HippoRAG 2 & $233.7$ & $38.9$ & $57$ & $9,670$ & $10,693$\\
    Raptor & $29.8$ & $10.0$ & $30$ & $6,067$ & $1,017$\\
    \midrule
    VanillaRAG & $2.9$ & $27.1$ & $0$ & $1,723$ & $418$\\
    \bottomrule
  \end{tabular}
\end{table}

We first analyze the time costs for both offline construction and online retrieval, with results summarized in Table~\ref{tbl:overall_cost}.
For the offline build time, the simpler VanillaRAG baseline is the fastest as expected due to its straightforward indexing process. Our proposed \systemname{} system requires $5.2$ hours to build the knowledge base on the $\text{\benchmarkname}_{M}$ corpus. While this is longer than the VanillaRAG baseline, it is significantly more efficient than other complex, graph-based frameworks like Raptor ($29.8$ hours). This demonstrates that our specialized pipeline achieves a favorable balance between the richness of the constructed KG and the computational cost of its construction. For the superior online performance it enables, this build process is cost-effective.
For the online retrieving time, which determines the user experience, \systemname{} is designed for high efficiency. Due to the pre-computed graph structure and the fast FAISS index, our system can achieve a low query latency of approximately $8.4$ seconds per query. This ensures a responsive and interactive search experience, which is a key requirement for any practical dataset discovery tool.

In addition to time, the LLM API token consumption is a critical factor for practical deployment. Table~\ref{tbl:overall_cost} details the token costs for both the offline building phase and the online query phase.
During the building phase, our multi-agent extraction module proves to be highly cost-effective. The Preliminary Filtering Agent discards irrelevant documents to significantly reduce the number of expensive calls to the more powerful Analyst and Enrichment Agents. Meanwhile, our hybrid dataset resolution method reduces repetitive LLM calls. These result in a substantially lower total token consumption compared to baseline methods that might perform deep processing on the entire corpus.
For the query phase, the primary API cost comes from the final LLM rerank step. As this step only processes a small number of top candidates, the per-query token cost is minimal. This design ensures that users can benefit from the high precision of LLM-based reranking without excessive running costs.

Finally, we analyze the storage footprint of the generated artifacts, as shown in Table~\ref{tbl:overall_cost}. The results indicate that our \systemname{} system maintains a moderate storage footprint. This surprising result comes from a fundamental architectural difference. A standard RAG approach must create vector embeddings for every text chunk across the entire corpus. In contrast, our \systemname{} pipeline first intelligently identifies and extracts the most relevant information. Consequently, our system only needs to embed and index this smaller and highly relevant subset of text. This strategy allows \systemname{} to build a rich and targeted knowledge base while maintaining a compact storage footprint.

In summary, our efficiency and cost analysis demonstrates that \systemname{} is a practical system. It achieves an effective balance of controlling reasonable one-time offline costs in terms of time, API calls, and storage to provide an efficient and cost-effective online query feature.

\begin{figure}[ht]
  \centering
  \includegraphics[width=0.93\linewidth]{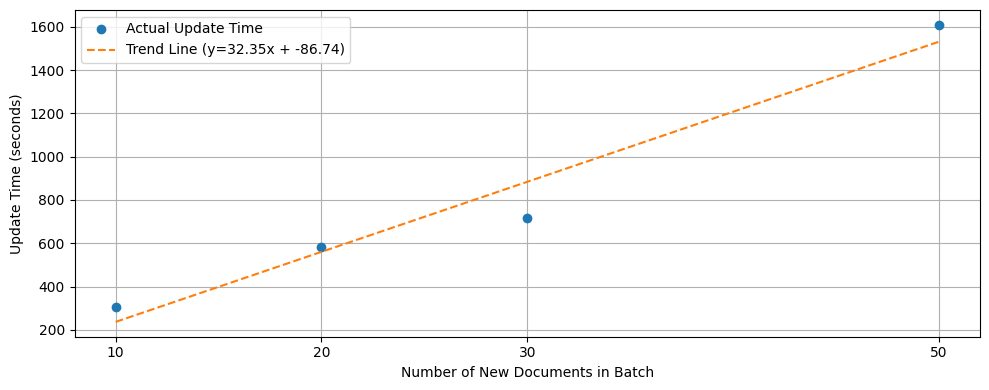}
  \caption{\revised{
   The number of new documents vs. the time cost of the incremental update, along with the fitted trend line.
  }}
  \label{fig:incremental_efficiency}
\end{figure}

\revised{
\subsection{Incremental Update Efficiency}

A crucial requirement for our system is the ability to efficiently incorporate new knowledge over time. To evaluate the efficiency and scalability of our incremental update pipeline, we conducted an experiment on the $\text{\benchmarkname}_{M}$ corpus. First, an initial knowledge base was built using a base set of 518 documents. Subsequently, we simulated the arrival of new data by adding the remaining documents in batches of varying sizes, and measured the time required for each update operation.

As shown in Figure~\ref{fig:incremental_efficiency}, there is a near-linear relationship between the number of new documents and the time required for the update process. The clear trend line indicates that the cost of an update is primarily dependent on the size of the incoming data. This confirms that our incremental approach is highly efficient and scalable, avoiding the prohibitive costs of performing a full system rebuild as the corpus grows.
}

\subsection{Ablation Studies}

To understand the individual contribution of key components in our \systemname{} system, we conduct a series of ablation studies. In these experiments, we systematically disable specific modules from our full pipeline and observe the impact on the end-to-end effectiveness. This allows us to quantify the value brought by each of our system design choices.

\begin{figure}[ht]
  \centering
  \includegraphics[width=0.87\linewidth]{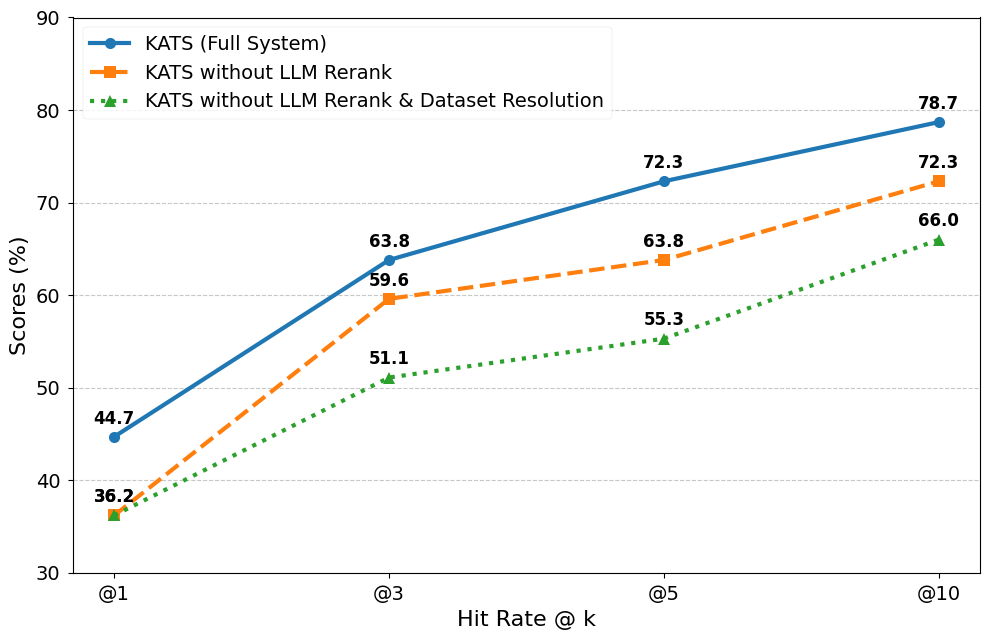}
  \caption{Results of ablation experiments on $\text{\benchmarkname}_{M}$.}
  \label{fig:ablation-all}
\end{figure}

\subsubsection{LLM Rerank}

We first investigate the impact of the final LLM-based reranking step. The blue and orange folds in Figure~\ref{fig:ablation-all} compare the performance of our full system with a variant where the reranking module is disabled. As illustrated, the inclusion of the LLM reranker brings a significant improvement in each level of hit rate. This demonstrates that while our graph-based ranking is effective at producing a strong set of candidates, the LLM's reasoning capabilities are crucial for fine-tuning the precision of the top-ranked results.

\subsubsection{Dataset Resolution}

Next, we evaluate the contribution of our dataset resolution module, which is designed to combat entity ambiguity. To better demonstrate the effect of this module on the KG, instead of using LLM reranking, we directly evaluate the search results from the KG in our experiments. As shown in the orange and green folds in Figure~\ref{fig:ablation-all}, disabling this module has little effect on top-1 hit rates, but results in significant performance degradation of other hit rate metrics. This result strongly validates our hypothesis that resolving dataset naming ambiguities is a critical factor in real-world dataset search. By merging duplicate entities, our system is able to integrate fragmented information to form a more complete and accurate KG, thus directly improving recall and relevance.

\subsection{Microbenchmarks}

In addition to end-to-end performance, we conduct a series of microbenchmarks to analyze the robustness and sensitivity of our \systemname{} system. Specifically, we investigate the system's performance concerning its key hyperparameter and its dependency on the underlying LLM.

\subsubsection{Hyperparameter Sensitivity}

\begin{figure}[ht]
  \centering
  \includegraphics[width=0.86\linewidth]{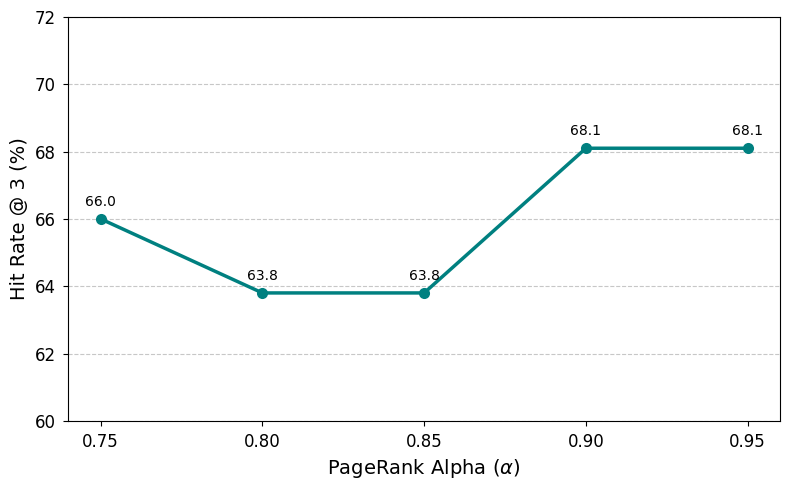}
  \caption{Sensitivity analysis of Hit Rate@3 with respect to the PageRank alpha ($\alpha$) parameter on $\text{\benchmarkname}_{M}$.}
  \label{fig:sensitivity-alpha}
\end{figure}

We first analyze the system's sensitivity to the choice of its key hyperparameter to ensure that its strong performance is not confined to a narrow setting. We focus on the PageRank alpha ($\alpha$) that balances the influence of the initial semantic search and the subsequent graph traversal.

For this purpose, we fix all other parameters and run our system on the $\text{\benchmarkname}_{M}$ corpus while varying the value of $\alpha$: $$\alpha \in \{0.75,0.80,0.85,0.90,0.95\}$$
The impact on the Hit Rate@3 metric is recorded.

As shown in Figure~\ref{fig:sensitivity-alpha}, the system maintains a high level of effectiveness across a reasonable range of $\alpha$ values. A lower alpha value, which puts more emphasis on graph exploration, leads to a slight degradation. This suggests that the initial semantic signal from the seed tasks is highly valuable in the experiment. However, as the hyperparameter changes, the overall performance remains relatively stable. This result indicates that \systemname{} is robust and not critically sensitive to the precise tuning of the hyperparameter $\alpha$.

\subsubsection{Robustness to LLMs}

\begin{table}[h]
  \centering
  \caption{\revised{Comparison of Hit Rate of \systemname{} with different LLMs for building and QA on $\text{\benchmarkname}_{M}$.}}
  \label{tbl:llm_robustness}
  \begin{tabular}{ll cccc}
    \toprule
    \textbf{Building} & \textbf{QA} & \textbf{@1} & \textbf{@3} & \textbf{@5} & \textbf{@10} \\
    \midrule    
    gpt-4o-mini & gpt-4o-mini & 0.447 & 0.638 & 0.723 & 0.787 \\
    gemini-1.5-flash & gpt-4o-mini & 0.426 & 0.690 & 0.746 & 0.787 \\
    gemini-1.5-flash & gemini-1.5-flash & 0.404 & 0.532 & 0.617 & 0.702 \\
    \bottomrule
  \end{tabular}
\end{table}

To verify that the effectiveness of our framework is not dependent on a specific proprietary model, we evaluate its robustness using a different LLM. Specifically, for building and QA phases, we replaced the gpt-4o-mini~\cite{openai2024gpt4omini} with a comparable model, gemini-1.5-flash~\cite{google2025gemini1_5flash}, and reran the end-to-end evaluation on the $\text{\benchmarkname}_{M}$ corpus.

\revised{
Table~\ref{tbl:llm_robustness} presents the results of this sensitivity analysis. The findings are twofold. Firstly, when only the LLM in the \textit{building} phase is substituted, \systemname's end-to-end performance remains remarkably stable, with no significant degradation in Hit Rate metrics. This strongly suggests that our pipeline is not sensitive to the specific choice of the extraction model. 

Secondly, this stability indicates that the downstream components of our architecture, i.e., the hybrid entity resolution and graph-based ranking, are robust enough to handle the potential variations and minor errors introduced by different extraction models. While using a weaker model for the QA phase predictably impacts performance, the system still outperforms the baselines. This experiment validates that the core strength of \systemname{} lies in its specialized architecture and strategy, rather than a dependency on any single proprietary model for its knowledge extraction process.
}

\section{Case Studies}
\label{sec:casestudies}

\revised{
\subsection{Incremental Knowledge Discovery}

To illustrate the practical utility of the incremental update feature, we present a case study demonstrating how \systemname{} can dynamically incorporate new knowledge to answer previously unanswerable queries.

In the experiment, a specific paper from the $\text{\benchmarkname}_{M}$ corpus that mentions the \textbf{CARLA Simulator Dataset} was deliberately withheld during the initial build process. Then, a query related to the dataset was issued as shown in Table~\ref{tbl:case_study_incremental}. Since the relevant document was absent, the system was unable to find and recommend the CARLA Simulator Dataset.

Afterwards, the withheld paper was introduced into the corpus through the incremental update process. The pipeline correctly and efficiently integrates the newly extracted task and dataset information into the existing knowledge base. When the exact same query was issued, \systemname{} successfully identified the \textbf{CARLA Simulator Dataset} as a relevant result for the task.

\begin{table}[htb]
    \centering
    \caption{\revised{Comparison of results for the incremental knowledge discovery case study.}}
    \label{tbl:case_study_incremental}
    \makebox[\columnwidth][c]{
        \begin{tabular}{l p{6cm}}
            \toprule[1pt]
            \multicolumn{2}{{p{\columnwidth}}}{\textbf{Query:} \textit{``Training models for urban driving scenarios using high-resolution visual inputs and expert actions.''}} \\
            \midrule[0.5pt]
            \textbf{Before Update} & Unable to find the CARLA Simulator Dataset.\\ \textbf{After Update} &
            Recommend CARLA as the top-2 result. \\
            \bottomrule[1pt]
        \end{tabular}
    }
\end{table}

This real-world case study demonstrates that the incremental update mechanism is efficient and effective. It empowers \systemname{} to evolve by integrating new documents, thereby keeping its dataset recommendations current as new research emerges.
}

\revised{
\subsection{Comparison with Online Research Agents}

A relevant point of comparison for \systemname{} is the emerging paradigm of online AI research agents, such as Gemini's Deep Research feature. 
While both approaches aim to answer complex queries, they represent fundamentally different technical routes. We will illustrate these differences through the following case studies, as shown in Table~\ref{tbl:case_study_online_agent}.
It is worth noting that this comparison inherently favors Deep Research, since the paper containing the ground-truth pair is hidden from KATS but remains accessible to the web-based agent.

\begin{table}[h!]
\centering
\caption{\revised{Performance comparison of the online research agent case study.}}
\label{tbl:case_study_online_agent}
\begin{tabular}{l ccc}
\toprule
\textbf{Task} & \textbf{Method} & \textbf{Latency} & \textbf{Result}\\
\midrule
\multirow{2}{*}{Case 1} & Gemini & $392$s & Ground truth within 6 answers \\
 & \systemname{} & $< 20$s & Ground truth ranks 5th \\
\midrule
\multirow{2}{*}{Case 2} & Gemini & ~$400$s & Ground truth within 6-8 answers \\
 & \systemname{} & $< 20$s & Ground truth ranks 3rd \\
\bottomrule
\end{tabular}
\end{table}

\textbf{Case 1: Physical Commonsense Reasoning Task}. Gemini Deep Research visited 111 websites, including some non-directional, unrelated websites such as Hugging Face Daily Papers. After $392$ seconds, Gemini provided $6$ answers, among which was the ground truth ``Kodak Dataset''. \systemname{} provided the answer within $20$ seconds, where the ground truth is the top-5 answer.

\textbf{Case 2: Image Compression Evaluation Task}. In $2$ attempts, Gemini Deep Research visited about $100$ websites, and within about $400$ seconds, Gemini separately provided $6$ and $8$ answers, among which was the ground truth ``PiQA''. \systemname{} provided the answer within $20$ seconds, where the ground truth is the top-3 answer, and this result remained consistent across multiple attempts.

From the two cases above, it is evident that \systemname{} and online research agent follow different paradigms. \systemname{} invests an offline cost to process a user-defined, domain-specific corpus into a persistent knowledge base. This approach yields significant advantages in \textit{query latency}, \textit{cost-efficiency}, and \textit{reproducibility}. Furthermore, its offline capability makes it ideal for private document collections, and the incremental update mechanism allows it to efficiently integrate new knowledge. In contrast, online research agents dynamically browse the general web for each query. This provides real-time information, but results in higher latency, greater cost, and less reproducible results sourced from a mix of professional and general-interest websites.
}

\revised{
\subsection{Refined Task Disambiguation}

A significant challenge in automated knowledge extraction arises from complex research papers that involve multiple evaluation tasks for distinct capabilities using different datasets. A naive extraction approach might incorrectly associate all datasets with a single broad task, which can reduce the precision of the KG. 

To demonstrate \systemname{}'s robustness against this challenge, we present a case study. From $\text{\benchmarkname}_{L}$, we select a paper that trains and evaluates an LLM across three disparate domains: mathematical reasoning, natural language understanding, and code generation. For these $3$ specific tasks, the paper uses GSM8K~\cite{cobbe2021gsm8k}, GLUE~\cite{wang2019glue}, and HumanEval~\cite{chen2021humaneval}, respectively.

Our multi-agent extraction framework successfully disambiguated these contexts. Instead of extracting a generic task like ``evaluate a model'', \systemname{} identified and linked specific tasks to their corresponding datasets, as shown in Table~\ref{tbl:case_study_extraction}.

\begin{table}[h]
  \centering
  \caption{\revised{Results of the task disambiguation case study.}}
  \label{tbl:case_study_extraction}
  \begin{tabularx}{\linewidth}{l X}
    \toprule
    \textbf{Dataset} & \textbf{Extracted Task Description} \\
    \midrule
    GSM8K     & \textit{``[...] multiple natural language understanding tasks''} \\
    GLUE      & \textit{``[...] multi-step mathematical word problems''} \\
    HumanEval & \textit{``[...] generate correct code solutions for programming tasks''} \\
    \bottomrule
  \end{tabularx}
\end{table}

More importantly, the resulting KG correctly represents these as three independent task-dataset branches, without creating erroneous links between them. Therefore, when querying for one specific task, the PPR algorithm does not assign high scores to the other two tasks, ensuring that the results returned by \systemname{} are precise. This case study empirically validates that \systemname{}'s architecture is capable of capturing context-specific relationships within a single document, and confirms the high precision of our automated KG construction process.
}

\section{Discussion}
\label{sec:discussion}

\subsection{Findings and Implications}

Our extensive experimental evaluation demonstrates that \systemname{}, a system specifically designed for task-oriented dataset search from unstructured text, significantly outperforms state-of-the-art general-purpose RAG frameworks in both effectiveness and efficiency. This finding leads to several key implications for the future of dataset discovery.

\subsubsection{From Keywords to Intent: Bridging the Semantic Gap}

A primary challenge highlighted by prior work is the significant ``semantic gap'' between a user's task-oriented need and the keyword-based queries that traditional systems require~\cite{Hulsebos2024ItTLMainSurvey}. Our work demonstrates a successful paradigm that bridges this gap. By accepting open-ended natural language descriptions as input, \systemname{} moves the cognitive burden of ``query formulation'' from the user to the system, which interprets complex task descriptions and map them to relevant datasets. Future data discovery systems should focus less on optimizing keyword matching, but more on developing models that can analyze complex user intent.

\subsubsection{From General-purpose to Specialized RAG}

Our results also suggest that while general-purpose RAG architectures provide powerful and flexible frameworks, significant performance gains can be achieved by designing specialized systems for a specific problem. For task-oriented dataset search, this specialization involves not only retrieval but also a deep integration of entity ambiguity and task-to-dataset mapping into the core pipeline. Developing specialized systems for high-value, vertical domains is a viable direction for RAG research.

\subsection{Limitations}

While our proposed \systemname{} system demonstrates significant effectiveness for task-oriented dataset search, we also recognize several limitations to be addressed in the future.

\subsubsection{Domain Specificity of Knowledge Source}

Our current pipeline, particularly the multi-agent extraction framework, is tailored for the unique characteristics of scientific literature. A key premise of our approach is that these documents typically co-locate the mention of a dataset with a rich description of the specific task for which it was applied. However, the effectiveness of our extraction strategy might vary when applied to other types of documents where this premise does not hold. Even though we have extended the system to integrate dataset descriptions without specific tasks, such as a dataset repository, the extraction process still requires a significant re-evaluation for other types of documents.

\subsubsection{Dependency on the Fidelity of LLM}

\revised{We acknowledge that, like many contemporary automated knowledge extraction systems \cite{edge2024graphrag,jimenez2025hipporag2,sarthi2024raptor}, the quality of our KG is fundamentally dependent on the fidelity of the upstream LLM used in the extraction and entity resolution stages. 
Possible errors from the LLM, such as an incorrect extraction, can propagate through the pipeline and 
lead to erroneous nodes or edges
in the graph. 
This dependency presents a valid concern. However, \systemname{} was designed with specific strategies to mitigate these risks. Firstly, our multi-agent extraction framework decomposes the complex end-to-end extraction task into a series of smaller, more manageable sub-tasks. By assigning focused responsibilities to different agents, we reduce the cognitive load on the LLM in any single step, minimizing the likelihood of errors such as hallucinations or overlooking details. Secondly, in critical stages like entity resolution, we deliberately constrain the role of the LLM. Instead of tasking the LLM with the open-ended challenge of clustering ambiguous entities, we employ a hybrid approach. The algorithmic methods first generate high-probability candidate pairs, and then the LLM performs a much simpler and more reliable binary verification task. 

The evidence presented in Table~\ref{tbl:llm_robustness}, which shows stable end-to-end performance when substituting the building phase LLM, proves the effectiveness of these mitigation strategies. Nevertheless, the current system still lacks a specialized module for automatically verifying the authenticity of all LLM outputs and correcting possible errors based on the source documents, which remains a limitation.
}

\section{Conclusion}
\label{sec:conclusion}

Task-oriented search of datasets is a critical but unmet need for data professionals, encountering the challenges of a persistent semantic gap, lack of task-to-dataset mapping and benchmarks, and entity ambiguity. 
To address these challenges, our work makes several key contributions. Firstly, we developed the \benchmarkname{} benchmark suite that addresses the critical gap in lacking evaluation benchmarks and enables the rigorous assessment of task-oriented dataset search systems. Moreover, we presented \systemname{}, a novel end-to-end pipeline that automatically constructs a high-quality KG offline with a dedicated entity resolution mechanism, to resolve the challenge of lacking task-to-dataset mapping and entity ambiguity. \revised{Importantly, we equipped this pipeline with an efficient incremental update mechanism, transforming the static knowledge base into a dynamic and scalable one.} To tackle the challenge of the persistent semantic gap, \systemname{} is equipped with an online query processing pipeline that integrates vector search with graph-based ranking to generate highly relevant and task-consistent dataset recommendations based on the KG constructed before. 

We believe that our work provides a solid foundation and a promising blueprint for the next generation of dataset discovery systems. \revised{Building on this foundation, there are several promising avenues for future work. A key direction is generalizing our framework beyond scientific literature to new problem domains, such as linking symptoms to diseases from medical texts. Further research could also focus on improving the technical capabilities, such as developing real-time streaming updates for the KG and incorporating automatic fact-checking modules to enhance its reliability.}

\begin{acks}
Chenhao Ma was partially supported by NSFC under Grant 62302421, Basic and Applied Basic Research Fund in Guangdong Province under Grant 2025A1515010439,  and the Guangdong Provincial Key Laboratory of Big Data Computing, The Chinese University of Hong Kong, Shenzhen. Xiaolin han was supported by NSFC under Grant 62302397, the China Postdoctoral Science Foundation under Grant Number 2025M774364, and Shaanxi Post-doctoral Research Project (2025BSHSDZZ106).
\end{acks}



\bibliographystyle{ACM-Reference-Format}
\bibliography{references}

@misc{pypdf,
  title = {pypdf},
  author = {{pypdf developers}},
  howpublished = {\url{https://github.com/py-pdf/pypdf}},
}

@misc{llamaparse,
  title = {LlamaParse},
  author = {{LlamaIndex Team}},
  year = {2025},
  howpublished = {\url{https://github.com/run-llama/llama_cloud_services}},
}

@inproceedings{jarrett2022online,
  title={Online Decision Mediation},
  author={Jarrett, Daniel and Hüyük, Alihan and van der Schaar, Mihaela},
  booktitle={Advances in Neural Information Processing Systems (NeurIPS)},
  year={2022}
}

@misc{openai2024embedding,
  author       = {OpenAI},
  title        = {text-embedding-3-small},
  year         = {2024},
  howpublished = {\url{https://platform.openai.com/docs/guides/embeddings/embedding-models}},
}

@article{douze2024faiss,
  title={The Faiss library},
  author={Matthijs Douze and Alexandr Krylov and Jeff Johnson and Herv{\'e} J{\'e}gou},
  journal={arXiv preprint arXiv:2401.08281},
  year={2024}
}

@article{edge2024graphrag,
  title={From Local to Global: A Graph RAG Approach to Query-Focused Summarization},
  author={Edge, Darren and Trinh, Ha and Cheng, Newman and Bradley, Joshua and Chao, Alex and Mody, Apurva and Truitt, Steven and Metropolitansky, Dasha and Ness, Robert Osazuwa and Larson, Jonathan},
  journal={arXiv preprint arXiv:2404.16130},
  year={2024}
}

@article{jimenez2025hipporag2,
  title={From RAG to Memory: Non-Parametric Continual Learning for Large Language Models},
  author={Jiménez Gutiérrez, Bernal and Shu, Yiheng and Qi, Weijian and Zhou, Sizhe and Su, Yu},
  journal={arXiv preprint arXiv:2502.14802},
  year={2025}
}

@inproceedings{sarthi2024raptor,
  title={RAPTOR: Recursive Abstractive Processing for Tree-Organized Retrieval},
  author={Sarthi, Parth and Abdullah, Salman and Tuli, Aditi and Khanna, Shubh and Goldie, Anna and Manning, Christopher D.},
  booktitle={International Conference on Learning Representations (ICLR)},
  year={2024},
}

@inproceedings{yang2018hotpotqa,
  title={HotpotQA: A Dataset for Diverse, Explainable Multi-hop Question Answering},
  author={Yang, Zhilin and Qi, Peng and Zhang, Saizheng and Bengio, Yoshua and Cohen, William W. and Salakhutdinov, Ruslan and Manning, Christopher D.},
  booktitle={Proceedings of the 2018 Conference on Empirical Methods in Natural Language Processing},
  pages={2369--2380},
  year={2018},
  organization={Association for Computational Linguistics},
  address={Brussels, Belgium},
  doi={10.18653/v1/D18-1259}
}

@inproceedings{thakur2021beir,
  title={BEIR: A Heterogenous Benchmark for Zero-shot Evaluation of Information Retrieval Models},
  author={Thakur, Nandan and Reimers, Nils and Rücklé, Andreas and Srivastava, Abhishek and Gurevych, Iryna},
  booktitle={Thirty-fifth Conference on Neural Information Processing Systems Datasets and Benchmarks Track (Round 2)},
  year={2021},
  url={https://arxiv.org/abs/2104.08663}
}

@inproceedings{Hulsebos2024ItTLMainSurvey,
title = {``It Took Longer than I was Expecting:'' Why is Dataset Search Still so Hard?},
author = {Madelon Hulsebos and Wenjing Lin and Shreya Shankar and Aditya G. Parameswaran},
booktitle = {Proceedings of the 2024 Workshop on Human-In-the-Loop Data Analytics (HILDA '24)},
year = {2024},
pages = {1--4},
address = {Santiago, Chile},
publisher = {ACM},
doi = {10.1145/3665939.3665959},
url = {https://dl.acm.org/doi/10.1145/3665939.3665959}
}

@article{neum-etal-2016JDIQ,
  author = {Neumaier, Sebastian and Umbrich, Jürgen and Polleres, Axel},
  title = {Automated Quality Assessment of Metadata across Open Data Portals},
  journal = {ACM Journal of Data and Information Quality (JDIQ)},
  volume = {8},
  number = {1},
  pages = {2},
  year = {2016},
  month = nov,
  doi = {10.1145/2964909},
  url = {http://polleres.net/publications/neum-etal-2016JDIQ.pdf},
}

@article{PatonChenWu2023,
  author = {Norman W. Paton and Jiaoyan Chen and Zhenyu Wu},
  title = {Dataset Discovery and Exploration: A Survey},
  journal = {ACM Computing Surveys},
  year = {2023},
  doi = {10.1145/3626521},
  url = {https://dl.acm.org/doi/10.1145/3626521},
  publisher = {ACM}
}

@inproceedings{BrickleyBurgessNoy2019googledatasetsearch,
  author = {Dan Brickley and Matthew Burgess and Natasha Noy},
  title = {Google Dataset Search: Building a search engine for datasets in an open Web ecosystem},
  booktitle = {Proceedings of the 2019 ACM Web Conference (WWW '19)},
  year = {2019},
  pages = {1365--1375},
  doi = {10.1145/3308558.3313685},
  publisher = {ACM},
  address = {New York, NY, USA},
  url = {https://dl.acm.org/doi/10.1145/3308558.3313685}
}

@inproceedings{Balaka2025Pneuma,
  title = {Pneuma: Leveraging LLMs for Tabular Data Representation and Retrieval in an End-to-End System},
  author = {Muhammad Imam Luthfi Balaka and David Alexander and Qiming Wang and Yue Gong and Adila Krisnadhi and Raul Castro Fernandez},
  booktitle = {Proceedings of the ACM SIGMOD International Conference on Management of Data},
  year = {2025},
  month = {June},
  doi = {10.1145/3725337},
  url = {https://arxiv.org/abs/2504.09207},
  note = {arXiv:2504.09207 [cs.DB]}
}

@inproceedings{Galhotra2023METAM,
  title={METAM: Goal-Oriented Data Discovery},
  author={Sainyam Galhotra and Yue Gong and Raul Castro Fernandez},
  booktitle={39th IEEE International Conference on Data Engineering (ICDE)},
  year={2023},
  month={April},
  address={Anaheim, CA, USA},
  note={arXiv:2304.09068 [cs.DB]},
  url={https://arxiv.org/abs/2304.09068},
  doi={10.48550/arXiv.2304.09068}
}

@inproceedings{Huang2024FastPrivate,
  title={The Fast and the Private: Task-based Dataset Search},
  author={Zezhou Huang and Jiaxiang Liu and Haonan Wang and Eugene Wu},
  booktitle={Conference on Innovative Data Systems Research (CIDR)},
  year={2024},
  month={January},
  note={arXiv:2308.05637 [cs.DB]},
  url={https://arxiv.org/abs/2308.05637},
  doi={10.48550/arXiv.2308.05637}
}

@inproceedings{jimenez2024hipporag,
  title={HippoRAG: Neurobiologically Inspired Long-Term Memory for Large Language Models},
  author={Bernal Jim{\'e}nez Guti{\'e}rrez and Yiheng Shu and Yu Gu and Michihiro Yasunaga and Yu Su},
  booktitle={Advances in Neural Information Processing Systems (NeurIPS) 37},
  year={2024},
  note={NeurIPS 2024},
  url={https://arxiv.org/abs/2405.14831},
}

@article{Chapman2019DatasetSurvey,
  author = {Adriane P. Chapman and Elena Simperl and Laura Koesten and George Konstantinidis and Luis Daniel Ibáñez and Emilia Kacprzak and Paul Groth},
  title = {Dataset search: a survey},
  journal = {The VLDB Journal},
  year = {2019},
  volume = {29},
  number = {1},
  pages = {251--272},
  doi = {10.1007/s00778-019-00564-x},
  url = {https://doi.org/10.1007/s00778-019-00564-x}
}

@inproceedings{lewis2020RAG,
  title={Retrieval-Augmented Generation for Knowledge-Intensive NLP Tasks},
  author={Lewis, Patrick and Perez, Ethan and Piktus, Aleksandra and Petroni, Fabio and Karpukhin, Vladimir and Goyal, Naman and Küttler, Heinrich and Lewis, Mike and Yih, Wen-tau and Rockt{\"a}schel, Tim and Riedel, Sebastian and Kiela, Douwe},
  booktitle={Advances in Neural Information Processing Systems 33 (NeurIPS 2020)},
  year={2020},
  url={https://proceedings.neurips.cc/paper/2020/hash/6b493230205f780e1bc26945df7481e5-Abstract.html},
}

@misc{zhao2023surveyLLM,
  title={A Survey of Large Language Models},
  author={Zhao, Wayne Xin and Zhou, Kun and Li, Junyi and Tang, Tianyi and Wang, Xiaolei and Hou, Yupeng and Min, Yingqian and Zhang, Beichen and Zhang, Junjie and Dong, Zican and Du, Yifan and Yang, Chen and Chen, Yushuo and Chen, Zhipeng and Jiang, Jinhao and Ren, Ruiyang and Li, Yifan and Tang, Xinyu and Liu, Zikang and Liu, Peiyu and Nie, Jian-Yun and Wen, Ji-Rong},
  year={2023},
  howpublished={\url{https://arxiv.org/abs/2303.18223}},
}

@misc{liu2025hoprag,
  title={HopRAG: Multi-Hop Reasoning for Logic-Aware Retrieval-Augmented Generation},
  author={Liu, Hao and Wang, Zhengren and Chen, Xi and Li, Zhiyu and Xiong, Feiyu and Yu, Qinhan and Zhang, Wentao},
  year={2025},
  eprint={2502.12442},
  archivePrefix={arXiv},
  primaryClass={cs.IR},
  note={arXiv preprint arXiv:2502.12442, version 2, last revised May 26, 2025},
  url={https://arxiv.org/abs/2502.12442}
}

@inproceedings{Bogatu2020DataLakeDatasetDiscovery,
  title={Dataset Discovery in Data Lakes},
  author={Alex Bogatu and Alvaro A. A. Fernandes and Norman W. Paton and Nikolaos Konstantinou},
  booktitle={Proceedings of the 36th IEEE International Conference on Data Engineering (ICDE)},
  year={2020},
  address={Dallas, United States},
  month={April},
  doi={10.1109/ICDE48307.2020.00067},
  url={https://arxiv.org/abs/2011.10427}
}

@inproceedings{rajan2018aurum,
  title={Aurum: A Data Discovery System},
  author={Rajan, Deepak and Franklin, Michael J. and Hacker, Stephan and Hellerstein, Joseph M. and Gonzalez, Joseph E.},
  booktitle={Proceedings of the 2018 International Conference on Management of Data (SIGMOD)},
  year={2018},
  pages={2109--2112},
  doi={10.1145/3183713.3196927},
  publisher={ACM}
}

@article{Nargesian2018TableUS,
  title={Table Union Search on Open Data},
  author={Fatemeh Nargesian and Erkang Zhu and Ken Q. Pu and Renée J. Miller},
  journal={Proceedings of the VLDB Endowment},
  volume={11},
  number={7},
  pages={813--825},
  year={2018},
  doi={10.14778/3192965.3192973},
  url={https://doi.org/10.14778/3192965.3192973}
}

@inproceedings{semexplorer,
author = {Wei, Zixin and Han, Jun and Han, Xiaolin and Ma, Chenhao},
title = {SemExplorer: A User Interface for Semantic Approach to Customized Dataset Search},
year = {2025},
isbn = {9798400715648},
publisher = {Association for Computing Machinery},
address = {New York, NY, USA},
url = {https://doi.org/10.1145/3722212.3725133},
doi = {10.1145/3722212.3725133},
booktitle = {Companion of the 2025 International Conference on Management of Data},
pages = {255–258},
numpages = {4},
location = {Berlin, Germany},
series = {SIGMOD/PODS '25}
}

@inproceedings{castelo2021auctus,
  title={Auctus: A Dataset Search Engine for Data Discovery and Augmentation},
  author={Castelo, Sonia and Rampin, R{\'e}mi and Santos, A{\'e}cio and Bessa, Aline and Chirigati, Fernando and Freire, Juliana},
  booktitle={Proceedings of the VLDB Endowment},
  volume={14},
  number={12},
  pages={2791--2794},
  year={2021},
  publisher={VLDB Endowment},
  doi={10.14778/3476311.3476346},
  url={https://doi.org/10.14778/3476311.3476346}
}

@Article{Hu2023AutoTUS,
 author = {Xuming Hu and Shen Wang and Xiao Qin and Chuan Lei and Zhengyuan Shen and Asterios Katsifodimos and Christos Faloutsos and George Karypis and Lijie Wen and Philip S. Yu},
 title = {Automatic table union search with tabular representation learning},
 year = {2023},
 url = {https://www.amazon.science/publications/automatic-table-union-search-with-tabular-representation-learning},
}

@misc{openai2024gpt4omini,
  author = {OpenAI},
  title = {{GPT-4o mini} [Large language model]},
  year = {2024},
  howpublished = {\url{https://platform.openai.com/docs/models/gpt-4o-mini}}
}

@misc{google2025gemini1_5flash,
  author = {Google},
  title = {{Gemini 1.5 Flash} [Multimodal large language model]},
  year = {2025},
  howpublished = {\url{https://cloud.google.com/vertex-ai/generative-ai/docs/models/gemini/1-5-flash}}
}

@inproceedings{personalizedpagerank,
author = {Jeh, Glen and Widom, Jennifer},
title = {Scaling personalized web search},
year = {2003},
isbn = {1581136803},
publisher = {Association for Computing Machinery},
address = {New York, NY, USA},
url = {https://doi.org/10.1145/775152.775191},
doi = {10.1145/775152.775191},
booktitle = {Proceedings of the 12th International Conference on World Wide Web},
pages = {271–279},
numpages = {9},
location = {Budapest, Hungary},
series = {WWW '03}
}

@inproceedings{karpukhin-etal-2020-dense,
  title = {Dense Passage Retrieval for Open-Domain Question Answering},
  author = {Vladimir Karpukhin and Barlas Oguz and Sewon Min and Patrick Lewis and Ledell Wu and Sergey Edunov and Danqi Chen and Wen-tau Yih},
  booktitle = {Proceedings of the 2020 Conference on Empirical Methods in Natural Language Processing (EMNLP)},
  year = {2020},
  address = {Online},
  publisher = {Association for Computational Linguistics},
  pages = {6769--6781},
  url = {https://aclanthology.org/2020.emnlp-main.550},
  doi = {10.18653/v1/2020.emnlp-main.550},
  month = nov
}

@techreport{cifar10,
  author = {Alex Krizhevsky},
  title = {Learning multiple layers of features from tiny images},
  institution = {},
  year = {2009}
}

@article{An2025LEDD,
  title={LEDD: Large Language Model-Empowered Data Discovery in Data Lakes},
  author={Qi An and Chihua Ying and Yuqing Zhu and Yihao Xu and Manwei Zhang and Jianmin Wang},
  journal={arXiv preprint arXiv:2502.15182},
  year={2025},
  url={https://arxiv.org/abs/2502.15182}
}

@misc{schema.org,
  author       = {{Schema.org Community}},
  title        = {Schema.org},
  howpublished = {\url{https://schema.org}},
  year         = {2011}
}

@article{akhtar2024croissant,
  title={Croissant: A Metadata Format for ML-Ready Datasets},
  author={Mubashara Akhtar and Omar Benjelloun and Costanza Conforti and Luca Foschini and Joan Giner-Miguelez and Pieter Gijsbers and Sujata Goswami and Nitisha Jain and Michalis Karamousadakis and Michael Kuchnik and Satyapriya Krishna and Sylvain Lesage and Quentin Lhoest and Pierre Marcenac and Manil Maskey and Peter Mattson and Luis Oala and Hamidah Oderinwale and Pierre Ruyssen and Tim Santos and Rajat Shinde and Elena Simperl and Arjun Suresh and Goeffry Thomas and Slava Tykhonov and Joaquin Vanschoren and Susheel Varma and Jos van der Velde and Steffen Vogler and Carole-Jean Wu and Luyao Zhang},
  journal={arXiv preprint arXiv:2403.19546},
  year={2024},
  note={Published at NeurIPS 2024 Datasets and Benchmark Track},
  url={https://arxiv.org/abs/2403.19546}
}

@article{zhou2025depth,
  title={In-depth Analysis of Graph-based RAG in a Unified Framework},
  author={Zhou, Yingli and Su, Yaodong and Sun, Youran and Wang, Shu and Wang, Taotao and He, Runyuan and Zhang, Yongwei and Liang, Sicong and Liu, Xilin and Ma, Yuchi and others},
  journal={arXiv preprint arXiv:2503.04338},
  year={2025}
}

@article{cobbe2021gsm8k,
  title={Training Verifiers to Solve Math Word Problems},
  author={Cobbe, Karl and Kosaraju, Vineet and Bavarian, Mohammad and Chen, Mark and Jun, Heewoo and Kaiser, Lukasz and Plappert, Matthias and Tworek, Jerry and Hilton, Jacob and Nakano, Reiichiro and Hesse, Christopher and Schulman, John},
  journal={arXiv preprint arXiv:2110.14168},
  year={2021}
}

@inproceedings{wang2019glue,
  title={GLUE: A Multi-Task Benchmark and Analysis Platform for Natural Language Understanding},
  author={Wang, Alex and Singh, Amanpreet and Michael, Julian and Hill, Felix and Levy, Omer and Bowman, Samuel R.},
  note={In the Proceedings of ICLR},
  year={2019}
}

@article{chen2021humaneval,
  title={Evaluating Large Language Models Trained on Code},
  author={Chen, Mark and Tworek, Jerry and Jun, Heewoo and Yuan, Qiming and Pinto, Henrique and Kaplan, Jared and Edwards, Harri and Burda, Yuri and Joseph, Nicholas and Agarwal, Sandhini and others},
  journal={arXiv preprint arXiv:2107.03374},
  year={2021}
}



\end{document}